\documentclass[sigconf]{acmart}
\AtBeginDocument{%
  \providecommand\BibTeX{{%
    \normalfont B\kern-0.5em{\scshape i\kern-0.25em b}\kern-0.8em\TeX}}}

\usepackage{enumitem}
\usepackage{tabularx}
\usepackage{array}
\usepackage{multirow}
\usepackage{soul}
\usepackage{xcolor}
\usepackage{colortbl}
\usepackage{calc}
\usepackage{pifont}
\usepackage{booktabs}

\definecolor{lightyellow}{RGB}{255,255,204}
\definecolor{lightblue}{RGB}{204,204,255}
\definecolor{lightgreen}{RGB}{204,255,204}
\definecolor{lightergreen}{rgb}{0.0, 0.8, 0.0}

\newcommand{\cmark}{\textcolor{lightergreen}{\ding{51}}}  
\newcommand{\xmark}{\textcolor{red}{\ding{55}}}  

{\ignorespacesafterend}

\setcopyright{rightsretained}
\copyrightyear{2025}
\acmYear{2025}
\acmDOI{10.1145/3706598.3713273}
\acmISBN{979-8-4007-1394-1/25/04}

\acmConference[CHI '25]{CHI Conference on Human Factors in Computing Systems}{April 26-May 1, 2025}{Yokohama, Japan}
\acmBooktitle{CHI Conference on Human Factors in Computing Systems (CHI '25), April 26-May 1, 2025, Yokohama, Japan}

\usepackage{etoolbox}
\makeatletter
\tracingpatches
\patchcmd{\maketitle}{\@copyrightpermission}{
   \begin{minipage}{0.3\columnwidth}
     \href{<https://creativecommons.org/licenses/by/4.0/>}{\includegraphics[width=0.90\textwidth]{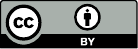}}
   \end{minipage}\hfill
   \begin{minipage}{0.7\columnwidth}
     \href{<https://creativecommons.org/licenses/by/4.0/>}{This work is licensed under a Creative Commons Attribution International 4.0 License.}
   \end{minipage}

   \vspace{5pt}
}{}{}

\makeatletter

\begin{document}

\title{No Evidence for LLMs Being Useful in Problem Reframing} 


\author{Joongi Shin}
\affiliation{%
  \institution{Aalto University}
  \city{Espoo}
  \country{Finland}}
\email{joongi.shin@aalto.fi}

\author{Anna Polyanskaya}
\affiliation{%
  \institution{Aalto University}
  \city{Espoo}
  \country{Finland}}
\email{apolyanskaya001@ikasle.ehu.eus}

\author{Andrés Lucero}
\affiliation{%
  \institution{Aalto University}
  \city{Espoo}
  \country{Finland}}
\email{lucero@acm.org}

\author{Antti Oulasvirta}
\affiliation{%
  \institution{Aalto University}
  \city{Espoo}
  \country{Finland}}
\email{antti.oulasvirta@aalto.fi}

\renewcommand{\shortauthors}{}
\newcommand{\rr}[1]{\textcolor{black}{#1}}
\newenvironment{rrSection}{\color{black}}{\ignorespacesafterend}

\begin{abstract}
Problem reframing is a designerly activity wherein alternative perspectives are created to recast what a stated design problem is about.
\rr{Generating alternative problem frames} is challenging because it requires devising novel and useful perspectives that fit the given problem context. Large language models (LLMs) \rr{could assist this activity via their generative capability}. 
However, it is not clear whether they can help designers produce high-quality frames. Therefore, we asked if there are benefits to working with LLMs. To this end, we compared three ways of using LLMs ($N=280$): 1) free-form, 2) direct generation, and 3) a structured approach informed by a theory of reframing. 
We found that using LLMs does not help improve the quality of problem frames. 
%
In fact, it increases the competence gap between experienced and inexperienced designers.
%
Also, inexperienced ones perceived lower agency when working with LLMs.
We conclude that there is no benefit to using LLMs in problem reframing and discuss possible factors for this lack of effect.

\end{abstract}





\maketitle


\section{Introduction}

\begin{figure*}[t]
    \centering
    \includegraphics[width=\linewidth]{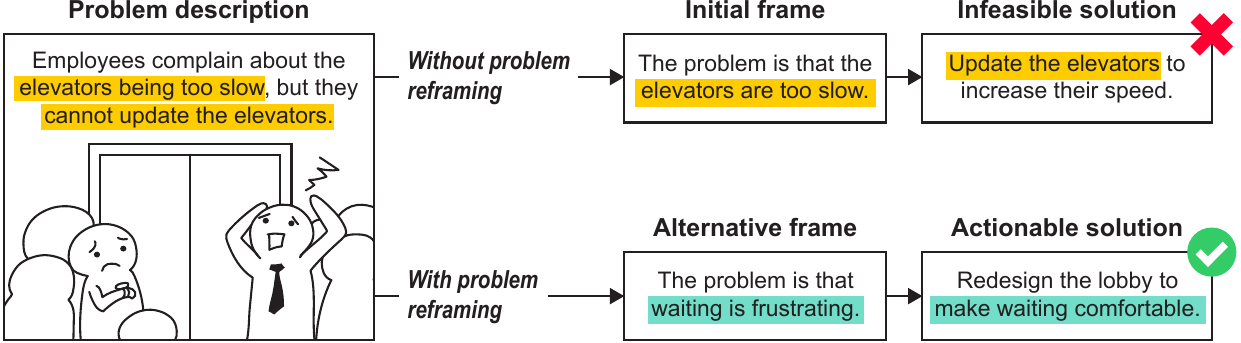}
    \caption{Problem reframing helps designers attack thorny design problems by rethinking what they are \emph{about}. 
    %
    In this paper, we study whether \rr{LLMs} can help designers arrive at better problem frames.}
    \Description{An illustration showing the concept of problem reframing. It shows an example problem and how it can lead to a limited solution without reframing or an actionable solution with reframing.}
    \label{fig:teaser}
\end{figure*}

\emph{Problem reframing} is a crucial designerly activity. 
By reframing a problem, a designer can make a challenging design problem more solvable \cite{dorst:book:reframing, schon:book:reflective, haase:2019:pf_meaningFrames, paton:2011:briefing_reframing}.
Therefore, a good problem frame for design is both novel and useful \rr{\cite{runco:2012:creativity}}.
Developing such frames requires exploring alternative ways to grapple with what a problem is \emph{about} \cite{dorst:co-evolve, grant:2011:pf_stakeholder_perspective}.
These points are illustrated by the ``slow elevator problem,'' outlined in Figure~\ref{fig:teaser}. Here, employees complain that elevators are too slow, while simply increasing the elevators' speed would cause safety issues.
By shifting perspective from seeing the problem as elevator speed (i.e., an initial frame) to the frustrating waiting experience (i.e., an alternative frame), designers could address employee complaints by improving the waiting area.
Without such reframings, fundamental problems might remain hidden, rendering the attempted solutions ineffective \cite{dorst:co-evolve}.

However, problem reframing is inherently challenging and labor-intensive, even for expert designers \cite{dorst:book:reframing}.
It demands putting great effort into understanding problems and altering perspectives \cite{dorst:book:reframing}. 
Designers need to iteratively ``destructure'' the initial problem, explore diverse stakeholder perspectives, synthesize overarching problem themes, and critically evaluate potential solutions \cite{dorst:framingProcess}.
Regrettably, such efforts can get caught in design fixation, which limits one's ability to explore problem frames \cite{dorst:co-evolve}. 
Failing in problem reframing may lead a designer to believe that the given problem is wicked or not solvable. 
%
%

With this paper, we look at \textit{whether working with AI -- specifically, large language models (LLMs) -- benefits designers in devising high-quality \rr{problem frames}} \footnote{We present an interactive demo of our project at \url{https://joongishin.github.io/problemReframing_llm/}}.
The question is timely: designers could be lured into using LLMs without understanding the effects on their creative potential. 
%
The use of LLMs in design activities has emerged as an important topic in the field of HCI \cite{shin:2023:human-ai, schmidt:2024:hcd-chatgpt, li:2024:interaction_pattern_LLM}.
Yet there are conflicting reports on how well LLMs can assist with creative activities.
On the one hand, prior studies have demonstrated LLMs' competence in generating various ideas that can inspire people to develop a wider range of ideas \cite{noy:2023:llm_writing_good, share:2024:brainwriting_llm, suh:2024:luminate, feng:2024:canvil_llm_ux}. The optimistic view has been that human--AI collaboration in creative activities lets practitioners focus on the core creative activities while delegating effortful tasks to AI \cite{li:2024:ux_perception_ai, schmidt:2024:hcd-chatgpt}.
On the other hand, some recent studies highlight LLMs' limitations in generating original ideas, thus implying that reliance on what LLMs generate can lead to less diversity of ideas \cite{doshi:2024:llm_writing_bad, anderson:2024:homogen_llm_ideation, koivisto:2023:human_outperform_ai}.
Despite the conflicting evidence, scholars have not given rigorous empirical attention specifically to LLM-assisted problem reframing. Research thus far has focused on uncovering LLMs' competence in generating alternative frames \cite{chen:2024:trizgpt, jiang:2024:autotriz, einarsson:2024:pf_ChatGPT, sandholm:2024:llm_randomness}, without considering their influence on designers. 
We argue that it is critical to consider designers and LLMs together when adressing this question. 
%
%
%


This paper contributes the first rigorously planned study to assess the benefits and drawbacks of using LLMs in problem reframing.
When planning the study, we wanted to capture both experiential and quality-related effects. 
Therefore, we focused on how using LLMs influences designers' creation of novel and original problem frames, alongside their perceptions of agency, ownership, and the helpfulness of LLMs.
Furthermore, we investigated how those influences might differ with designers' level of expertise in problem reframing.
In all, we tested four hypotheses about using LLMs in problem reframing (Section \ref{s:hypotheses}).

%

Reflecting the fact that LLMs can be used in a variety of ways beyond prompting, our research covered more than one approach to using them.
We included three distinctive ways of using LLMs found in the literature about creative activities \cite{anderson:2024:homogen_llm_ideation, suh:2024:luminate, xu:2024:llm_japmlate_why, jiang:2024:autotriz, he:2024:ai_groupIdeation}:
1) a \textsc{direct} approach wherein designers interact with LLM-generated problem frames only;
2) a \textsc{structured} one in which they reframe problems with LLMs in a step-by-step process, here following a nine-step process articulated by Kees Dorst \cite{dorst:book:reframing}); and
3) a \textsc{free-form} approach wherein designers converse with LLMs to reframe problems in their own strategy.
We tested each approach with \rr{OpenAI's GPT-4o}\footnote{ \url{https://platform.openai.com/docs/models/gpt-4o}}, the best-performing model publicly available \rr{at the time of our experiment (in August 2024)\footnote{\rr{\url{https://huggingface.co/spaces/ArtificialAnalysis/LLM-Performance-Leaderboard}} (note that GPT-o1 models were released in September 2024.)}}.
 

The study involves a large number of participants with design experience ($N = 280$).
The sample size was designed to achieve sufficient power for detecting medium effect sizes with high confidence. 
We created three challenging design problems and randomly assigned participants to one of the approaches and problems.
To aid in testing our hypotheses, we measured participants' expertise \rr{with a quiz that checked their conceptual and procedural understanding of problem reframing. In the end, participants with high expertise ($N = 15$) assessed the novelty and usefulness of generated frames. The participants' perceptions of using LLMs were measured with surveys \cite{doshi:2024:llm_writing_bad, chan:2022:positive_negative_agency, csi, hart:2006:nasa}.} 
The entire study was run on crowdsourcing platforms with careful quality controls in place.

Our main finding is that using LLMs in problem reframing does not aid designers in generating more novel or useful problem frames. The data show no statistically significant differences in quality between frames generated with and without LLMs.
In fact, we observed that using LLMs in problem reframing increased the \textit{competence gap} among designers. While frame quality did not differ in statistical terms across designers with different levels of understanding of problem reframing, using LLMs influenced more competent designers such that they created more novel frames than less competent ones did.
We observed a similar pattern in perceived agency, wherein more competent designers perceived higher levels than less competent designers.
We conclude the paper by discussing potential reasons for this lack of effect.
\rr{Our findings contribute to the understanding of human--AI interaction in design activities that could nurture designers' creative competence rather than merely replace them with AI.}
%

\section{Related Work}

Problem reframing is often misconstrued as rephrasing or improving a problem description, but that misses out on the core idea: exploring what the problem is about \cite{haase:2019:pf_meaningFrames, segal:2004:creativity_incubation, smith:1991:problem-solving_incubation}.
This section clarifies the concept of problem reframing and then reviews Dorst's structured process \rr{\cite{dorst:book:reframing}}. After that, we review prior work on the effects of using \rr{LLMs} in creative activities, which provides a foundation for our hypotheses.

\begin{figure*}[t]
    \centering
    \includegraphics[width=\linewidth]{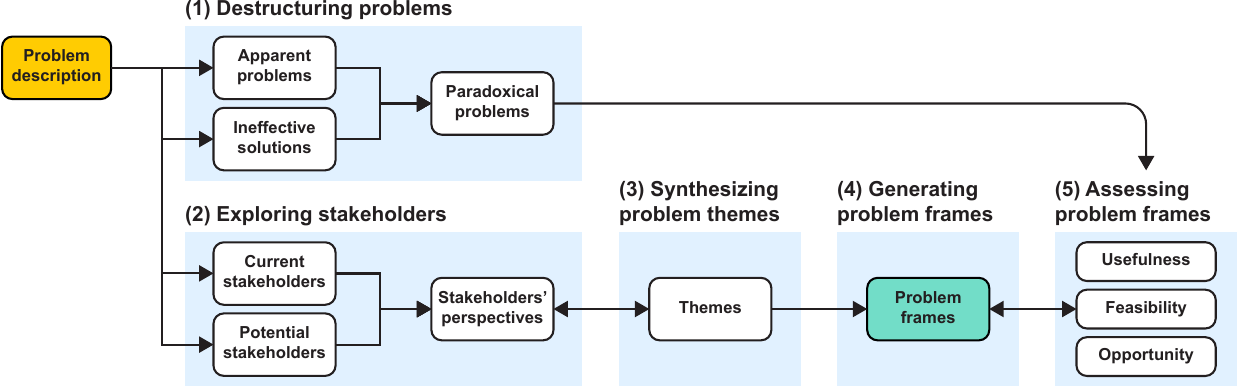}
    \caption{A structured problem-reframing workflow based on Dorst's reframing process \cite{dorst:book:reframing}.
    From the starting point of a description of problems (in the yellow box), intermediate content (in the white boxes) is generated, to lead to diverse problem frames (in the green box). \rr{We tested this flow as one of the approaches to LLM use in this study.}}
    \Description{A diagram showing the structured problem-reframing workflow based on Dorst's nine-step problem-reframing process.}
    \label{fig:dorst_process}
\end{figure*}

\subsection{Problem Reframing in Design}
To characterize framing, Donald Schön presents it as a reflective practice of sense-making \cite{schon:book:reflective}.
By actively redefining the nature of problems and potential solutions, designers deepen their understanding of the problematic situation and determine which factors to address and how \cite{schon:book:reflective, dorst:co-evolve}.
In this process, they employ multiple strategies to see ``the problem situation differently than before'' \cite{dorst:book:reframing}.
For example, designers analyze various stakeholders' viewpoints to uncover the underlying factors, and they use metaphors to reformulate problems outside of their initial contexts
\cite{silk:pf_cognitiveStyle, pee:2015:pf_seeing_different}.
They also engage in iterative co-evolution of the problem and solution spaces, exploring potential solutions in light of the current framing and redefining the problem on the basis of the solutions \cite{crilly:2021:co-evolution}.
In the end, their increased understanding of problems enables them to create new perspectives that fit the context of the problem \cite{dorst:co-evolve}.
In short, successful problem reframing relies on designers' understanding of problems that is enhanced by exploring diverse perspectives.
\rr{We focus on this reflective aspect that makes problem reframing a unique creative activity, as compared to merely diversifying solutions.}

\subsection{A Structured Process for Reframing}
Building on Schön's notion of problem framing, numerous studies have looked into how designers reframe problems in practice \cite{haase:2019:pf_meaningFrames, dorst:co-evolve, gao:pf_collaboration, silk:pf_cognitiveStyle, stompff:2016:pf_surprise}.
Dorst pioneered in articulating the problem-reframing process by means of in-depth analysis from multiple case studies \cite{dorst:book:reframing}.
According to him, designers can reframe problems in nine steps: 1) identifying apparent problems and previous, ineffective solutions; 2) defining what makes the problems difficult to solve; 3) investigating important current stakeholders and their perspectives; 4) expanding the problem context by identifying potential stakeholders; 5) identifying the underlying problem themes; 6) exploring in what ways the problem themes can be approached; 7) seeking assurance that the frames can lead to useful solutions; 8) evaluating the feasibility of solutions based on the frames; and 9) investigating new opportunities that alternative frames could bring, in addition to solving of the initial problems.
 
Relative to the other problem-reframing processes \cite{bardwell:1991:pf_process, Mensch:2019:pf_canvas, gao:pf_collaboration}, Dorst's descriptions provide a clearer workflow for applying the outcomes of each step as input to the next steps (see Figure \ref{fig:dorst_process}).
This aligns well with the input--output interaction style of LLMs,
\rr{where a complex task gets performed in a sequence of multiple prompts that take the output of previous prompts as the input (i.e., prompt chaining \cite{promptChaining:wu:2022}).}
Accordingly, we adopted Dorst's frame creation process to explore the use of LLMs in problem reframing.

\subsection{Conflicting Evidence on LLM-assisted Creative Activities}
Considering LLMs' ability to generate diverse ideas swiftly, numerous studies attest that they can contribute to creative activities \cite{qin:2024:ai_character_writing, lin:2024:ai_speech_writing, chakrabarty:2024:llm_writers, yuan:2022:wordcraft_writing, he:2024:ai_groupIdeation, llm_creativity:dis:2022, llm_creativity:dis:2023}.
Noy and Zhang \cite{noy:2023:llm_writing_good} showed that co-writing with LLMs can improve writers' productivity by 37\%.
Looking at the context of idea generation, Shaer et al. \cite{share:2024:brainwriting_llm} demonstrated direct prompting of LLMs to generate additional ideas for users to refine; they found that more than half of the ideas preferred by users combined human- and LLM-generated ideas. This suggests that LLMs offer utility for ideation tasks.
Taking a more structured approach, Suh et al. \cite{suh:2024:luminate} used LLMs to generate the design dimensions of poetry (e.g., moods or tones), then generate a spectrum of poetry along those dimensions. Their findings imply that using LLMs to support a structured ideation process meshes well with people's creation workflows for discovering more novel ideas.

Conversely, some recent studies suggest that such ways of using LLMs do not lead to people devising ideas that show greater creativity
\cite{ashkinaze:2024:human_ai_evolve_idea, chakrabarty:2024:llm_false_promise_art}.
For example, Doshi and Hauser's work connected with story-writing \cite{doshi:2024:llm_writing_bad} examined how LLM-generated ideas influence writers across different levels of creativity. While the authors found that exposing less creative writers to an LLM's ideas can significantly improve the novelty and usefulness of their stories (by 6.3\% and 10.7\%, respectively), it did not help more creative writers make significant improvements.
In another setting (which involved interacting with ChatGPT), Anderson et al. \cite{anderson:2024:homogen_llm_ideation} found that individuals generating ideas with LLMs produced ideas that were semantically similar to each other. This finding suggests that working with LLMs might actually limit the ideas' originality.
Similarly, Koivisto and Grassini \cite{koivisto:2023:human_outperform_ai} concluded that, while ideas from LLMs (here, ChatGPT with GPT-3.5 and -4) were more varied and creative on average, the best ideas still came from human agents.
In summary, while research points to potential for LLMs to enhance people's creative activities, LLMs' benefits and limitations here are not fully understood. We set out to tackle this knowledge gap by confirming whether they can assist in problem reframing.





\subsection{LLMs for Reframing Design Problems}
While LLMs' use in problem reframing has not been investigated sufficiently, a few studies point to their potential to help designers in the related subtasks.
Xu et al. \cite{xu:2024:llm_japmlate_why} demonstrated LLMs facilitating ``5-Whys'' design, a common activity for exploring the cause-and-effect underpinnings of a problem. For this, the authors implemented turn-based interaction wherein LLMs sequentially summarize users' speculation on causality, share their own perspectives, and generate a follow-up question to involve users in finding the root cause of a problem.
Also, there have been a few studies of using LLMs to facilitate a sense-making activity by generating abstractions of new concepts at multiple levels \cite{suh:2023:llm_sensemaking, gero:2024:llm_sensemaking}, which could be used to assist designers in destructuring problems.
%
Proceeding from such results, we hypothesized that LLMs' generative capabilities can be exploited to engage designers in a joint problem-reframing process.

For a wider context of problem-solving \cite{simon:1995:problemSolving, smith:1991:problem-solving_incubation}, researchers have hypothesized that training LLMs on previous successful solutions to diverse problems could equip them to generate useful solutions.
This idea has been tested specifically with the Theory of Inventive Problem Solving, or TRIZ method \cite{altshuller:1999:triz}, for solving problems by adapting the principles and patterns of solutions to similar problems in different domains.
For example, recent research has exploited LLMs to identify and point to existing solutions from relevant problem descriptions \cite{jiang:2024:autotriz, chen:2024:trizgpt}.
Sandholm et al. \cite{sandholm:2024:llm_randomness} demonstrated that LLMs fine-tuned via solution--problem datasets can generate problems and solutions paralleling the given problem. In their system, LLMs iterate between finding related solutions and problems, gradually diverging from the initial problem.
While promising, prior work focused exclusively on the LLMs' competence, by measuring the spread of the LLM-generated problems and solutions. 
\rr{Meanwhile, the influence of LLM use on designers' competence in deriving high-quality frames went without evaluation. Our study, in contrast, produced evidence of the benefits and limitations in LLMs' assistance to designers.}



\section{Hypotheses}
\label{s:hypotheses}


Our primary goal was to understand the influence of using LLMs in problem reframing. 
To this end, we first defined what constitutes LLMs' influence (e.g., frame quality and the designer's motivation in the reframing process) and what the key aspects are to consider in work with LLMs (e.g., designer expertise levels and varied ways of using LLMs).


\subsection{Effect on the Quality of Problem Frames}
While there is no established convention as to what constitutes ``good'' problem frames, previous work has reached some consensus in that framings with originality and that imply actionable solutions are preferred \cite{dorst:co-evolve, grant:2011:pf_stakeholder_perspective}.
Therefore, ideal assistance from LLMs should help designers create more novel and useful frames, not just make them review diverse perspectives.
Studies have shown that LLMs possess potential for this through their ability to supply diverse ideas \cite{llm_creativity:dis:2022, llm_creativity:dis:2023}. There is evidence that users can increase idea variety and quality by building on LLM-generated ideas \cite{ashkinaze:2024:human_ai_evolve_idea, liu:2022:prompt_image_chart}.
The overarching theory is that providing inspirational ideas can spark human association of seemingly unrelated concepts \cite{siangliulue:cscw:2015, goncalves:ijdci:2013, kim:dis:2006} and generating numerous ideas can lead to identifying more creative ones \cite{paulus:jcb:2011}.
In accordance with this standpoint, our primary hypothesis was this:
\begin{quote}
    \textbf{H1}: Using an LLM during problem reframing increases the \textit{novelty and usefulness} of the outputs.
\end{quote}

There are contrasting views: using LLMs might not help improve the problem frames' quality.
There are two possible reasons. Firstly, studies show that LLMs' ideas are less variety-rich and original than humans' \cite{koivisto:2023:human_outperform_ai, anderson:2024:homogen_llm_ideation, doshi:2024:llm_writing_bad}; hence, building on these could lead to low-quality problem frames. Secondly, evidence indicates that LLM-generated text and images are mostly high fidelity, which can cause users to fixate on them rather than explore alternative ideas \cite{wadi:2024:Genai_creativity_fixation}.


\subsection{Effect on Felt Agency and Ownership}
Secondly, we ask how using LLMs influences designers' sense of agency in the problem-reframing process and ownership in generated frames.
Felt agency refers to the perception of controlling one's own actions \cite{wen:2022:sense_of_agency}, and ownership is defined as the extent to which individuals feel that they own their creation \cite{zheng:2018:ownership_bias}.
%
\rr{Understanding effects on such perceptions can help promote individuals' motivation in creative activities \cite{amabile:1983:creativity_motivation, kroper:2011:motivation_design_creativity, li:2024:motivation_empathy_creativity}, which is particularly important in problem reframing, since the designers must shape their reframing process \cite{schon:book:reflective}.}
Dorst and Cross \cite{dorst:co-evolve} showed that designers actively seek additional information about problem situations when they find their current understanding insufficient for generating innovative problem frames. This need-driven exploration lets designers discover opportunities for reframing, whereby moments of creative insight or pleasant ``surprise'' unfold \cite{stompff:2016:pf_surprise}.
 
\rr{If using LLMs reduces designers' perceived agency and ownership, it could diminish their incentive for further exploration.}
As noted above, prior research presents a mixed view. While a few studies suggest that individuals may perceive less agency when they build upon AI-generated ideas \cite{moruzzi:2022:AI_agency, chan:2022:positive_negative_agency, lawton:2023:co-drawing_agency}, other work shows that users can retain a strong sense of agency when they are guided to iteratively refine AI-generated content \cite{long:2024:novelty_ai_workflow}.
\rr{The nature of any influence in the context of problem reframing remained unknown.}
%
%
We formed our second hypothesis accordingly:
\begin{quote}
    \textbf{H2}: Using an LLM during problem reframing decreases \textit{felt agency and ownership}.
\end{quote}


\subsection{Effect of the Way in Which LLMs are Used}

\begin{figure*}[t]
    \centering
    \includegraphics[width=\linewidth]{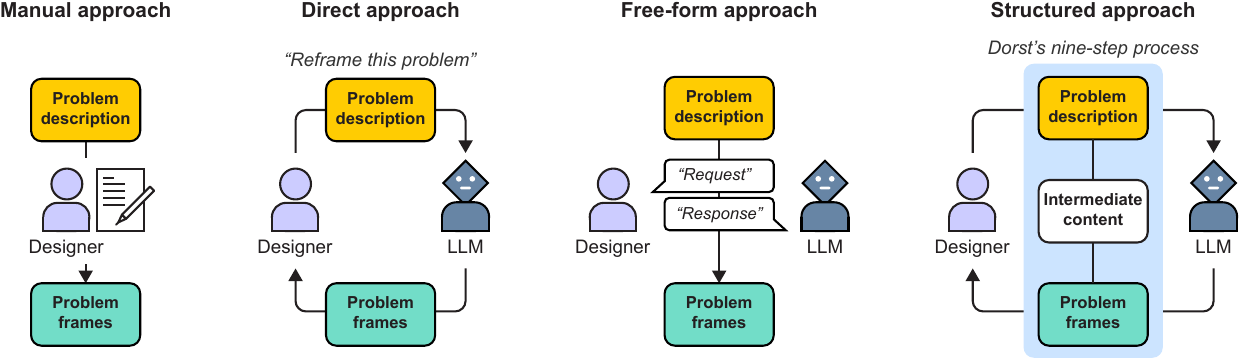}
    \caption{Compared to the \textsc{manual} approach where designers reframe problems by themselves (left),
    We tested three potential ways of using LLMs in problem reframing found in the literature about co-creation with LLMs. In the \textsc{direct} approach, designers can build on LLM-generated problem frames only. In the \textsc{free-form} approach, designers can freely ask LLMs to perform the tasks they require in their own process. In the \textsc{structured} approach, designers can also build on the intermediate content that LLMs generate following Dorst's nine-step reframing process \cite{dorst:book:reframing}.}
    \Description{A conceptual diagram showing how each problem-reframing approach derives alternative problem frames from problem description.}
    \label{fig:approaches}
\end{figure*}

We investigate how particular ways of using LLMs differently influence problem reframing.
Building on the prior research into using LLMs in creative activities \cite{anderson:2024:homogen_llm_ideation, suh:2024:luminate, xu:2024:llm_japmlate_why}, we looked at three distinctive methods, as shown in Figure \ref{fig:approaches}:

\begin{itemize}
    \item In the \textsc{direct} approach, designers prompt LLMs to directly generate problem frames. The LLM takes problem descriptions and generates alternative frames. How the designers build on the frames depends entirely on their expertise and preference.
    \item  The \textsc{free-form} approach allows designers to freely converse with LLMs. Here, designers can ask LLMs to perform various tasks as needed throughout their reframing process.
    \item Under the \textsc{structured} approach, the designer uses LLMs in a structured reframing process (Dorst's nine-step process \cite{dorst:book:reframing}). LLMs generate intermediate content (see the white boxes in Figure 2) and alternative frames. This allows building from the LLMs' outputs earlier on, instead of just with regard to final frames. 
\end{itemize}


While both \textsc{direct} and \textsc{free-form} approaches have unique benefits, we expect the \textsc{structured} approach to offer a more guided experience.
\rr{By inferring key information (e.g., why the problem is difficult to solve and who the potential stakeholders are) from a problem description, the \textsc{structured} approach could help designers spot alternative perspectives on the initial problem (Appendix \ref{appendix:system_structure} details the steps for this).}
According to theories on creative activities, people can be more creative when applying structured ideation methods \cite{sturdee:2015:structured_ideation, mora:2017:structured_ideation_tool}.
Also, studies on human--AI collaborative systems attest to the benefits of using LLMs in a structured manner \cite{suh:2024:luminate, xu:2024:llm_japmlate_why}. They help users to understand the process and ideate more systematically.
Therefore, our third hypothesis was this:
\begin{quote}
    \textbf{H3}: Guiding designers to use an LLM in a structured manner increases their \textit{perceived helpfulness} of LLMs.
\end{quote}


\subsection{Effect of Expertise}
We examined how designers' expertise influences reframing of problems with LLMs.
In the context of human-AI collaboration, research proves that people facing complex tasks, such as problem reframing, often lean on AI-generated suggestions \cite{fogliato:facct:2022, passi:2022:ai_overreliance, klingbeil:2024:ai_overreliance}.
This reliance is particularly noticeable among individuals with low confidence in their creative abilities \cite{ashkinaze:2024:human_ai_evolve_idea}, while LLM-generated ideas exert less influence on more creative people \cite{doshi:2024:llm_writing_bad}.

With these givens, we hypothesized that level of design expertise significantly influences how designers utilize LLM-generated problem frames.
%
While systematic exploration poses a challenge even for expert designers, research suggests that knowledge of the concept and process of problem reframing can help them conduct it \cite{gao:pf_collaboration, dorst:book:reframing}.
In contrast, designers who are not familiar with the process may struggle, tackling the problems as stated and evaluating the frames from narrower angles \cite{silk:pf_cognitiveStyle, cross:2004:design_expertise}.
Therefore, we expected designers less knowledgable of problem reframing \rr{(novices)} to be more likely to adopt LLM-generated problem frames, which might render them more vulnerable when LLMs generate low-quality frames.
In contrast, more knowledgeable designers \rr{(experts)} may rely on their expertise, filtering out generic perspectives from the LLM output or using them as seeds for original ones.
Such behaviors could also influence their experience of agency in the process, with expert designers retaining more felt agency than novices.
Therefore, this was our fourth hypothesis:
\begin{quote}
    \textbf{H4}: Using an LLM benefits designers with higher level of \textit{expertise in problem reframing} more.
\end{quote}
\section{Method}

\begin{figure*}[t]
    \centering
    \includegraphics[width=\linewidth]{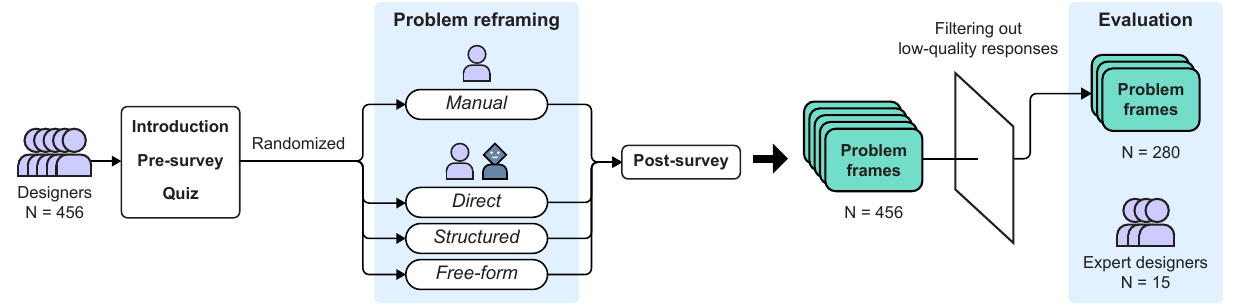}
    \caption{We conducted an empirical study with 456 participants. Prior to the main task, all participants went through the same introduction, pre-survey, and quiz. Each was randomly assigned an approach and instructed to reframe a problem. The study concluded with a post-survey. We collected 456 problem frames and filtered out the low-quality responses. In phase 2, 280 frames were evaluated, by 15 expert designers.}
    \Description{A diagram showing the structure of our study. From left to right, it shows a step-by-step process of completing our study. Sequentially, it shows the number of participants; the introduction phase with a pre-survey and quiz; randomly assigning participants to four conditions; the reframing phase; a post-survey; filtering out low-quality responses; and the evaluation phase with a final number of frame and expert designers.}
    \label{fig:study}
\end{figure*}

We tested these hypotheses in an online experiment with realistic tasks.
To guarantee a large enough sample for their robust testing, we handled the recruitment via crowdsourcing platforms\footnote{\rr{Prolific (\url{https://www.prolific.com/}) and CloudResearch (\url{https://www.cloudresearch.com/)}}}. 
The study had two phases: (1) reframing and (2) expert evaluation.
In the reframing phase, we measured participants' understanding of problem reframing and their perceived agency in working with LLMs.
Alongside the three LLM-based approaches -- \textsc{direct}, \textsc{structured}, and \textsc{free-form} --
we prepared a baseline condition wherein designers reframe problems without using LLMs (i.e., a \textsc{manual} setting).
We randomly assigned participants to one of the conditions, gave them a challenging design problem, and instructed them to generate two alternative frames they found most novel and useful.
For the evaluation by experts, we recruited designers with a strong understanding of problem reframing and had them assess the quality of problem frames generated in the first phase.
Our study design is shown in Figure \ref{fig:study}.

\subsection{Experiment Design}
Our study followed a between-subjects design with four levels: \rr{(\textsc{manual}, \textsc{direct}, \textsc{free-form}, and \textsc{structured})}.
Participants were assigned to these conditions in line with balanced Latin square randomization.
\rr{The conditions remained hidden from participants in both the reframing and the expert-evaluation phase, to prevent an influence on their behaviors; i.e., we used a blinded experiment design wherein participants did not know the condition from which the frames were generated or even of other conditions' existence.}

\subsection{Participants}
We recruited 456 individuals by means of the two crowdsourcing platforms.
For data quality, we filtered out those who failed at attention-check questions, submitted problem frames in incorrect format, or spent less than 10 minutes on the reframing task. We also removed people who were assigned to LLM-based approaches but did not use LLMs. In the end, we had 280 participants (145 identifying as women, 128 as men, and seven as ``other''; \rr{other demographics are presented in Appendix \ref{appendix:demographic}}), as shown in Figure \ref{fig:expertise}.

Participants were from diverse fields, such as product, UI/UX, graphic, and game design; architecture; and fashion. 
In total, 76\% had more than three years of design experience.
All but five participants had experience of reframing problems in design projects.
Of the latter, 43\% shared that they faced challenges in problem reframing, such as lack of information about the initial problems,  overcoming their biases and fixation on the problem's given form, and identifying stakeholders' fundamental needs.
More than half of the participants (56\%) had experience in using AI in the course of design tasks (ChatGPT, Midjourney, Adobe Firefly, etc).
From that subset, 83\% had an impression of AI as useful for their design tasks, while the rest perceived AI as not useful/sophisticated enough to support those tasks.
All participants were compensated with 16 euros for their hour of effort.

\subsection{Task and Materials}
\paragraph{Design problems.}
We designed the three design problems carefully to be challenging yet reframable. If the problems assigned are too easy to solve, most designers are bound to succeed without having to engage with LLMs. With overly difficult problems, though, participants might fail at reframing them within the time allocated.
Aware of these factors, we developed the problems around general social issues: aging, misbehavior, and preference conflict. These problems \rr{(described in Appendix \ref{appendix:problems})} are widely relatable and can be understood without the benefit of domain-specific knowledge. 
Via a pilot study (\emph{N}~=~36), we confirmed that the problems had an appropriate level of difficulty by examining participants' creation of a variety of low- and high-quality frames \rr{(see Appendix \ref{appendix:pilot})}.
Table \ref{tab:example_frame} presents one of the problems and some frames created for it.

\paragraph{LLM conditions.}
We designed prompts and interfaces to implement the \textsc{direct} and \textsc{structured} approaches \rr{(see Appendix \ref{appendix:system})}.
Our goal was to understand how designers can build on LLM-generated ideas under each approach. However, the quality of LLM outputs is heavily influenced by individuals' prompting skills \cite{zamfirescu:chi:2023}, which can affect the consistency of interaction across participants.
We found it crucial, therefore, to make sure the LLM-generated ideas' quality remained similar for all participants, enabling them to effectively reframe problems via both approaches.
Accordingly, we designed a standardized set of prompts by following the guidance of prompt-engineering literature ~\cite{white:arxiv:2023:prompt, Wynter:arxiv:2023, prompt/cox2023prompting, prompt/simmons2023moral} and iteratively experimenting with GPT-4o.
On our interface setting, participants (re)generate LLM output by pressing buttons, without needing to craft the prompts themselves.

\paragraph{Task instructions.}
In the reframing phase, \rr{participants' primary task} was to reframe a given problem, using the assigned approach. 
They were instructed to explore diverse perspectives on the problem and pinpoint the two problem frames that they found to be the most novel and useful.
\rr{In the LLM-based settings, participants were free to directly submit frames or build on LLM-generated ones, thereby presenting realistic use of LLMs.}
Additionally, they were asked to supply an explanation conveying how each of their frames could contribute to solving the problem. 
Then, in the post-survey, we requested them to select the frame they liked most and provide reasons.
This was to prevent participants from simply copying and pasting LLM-generated frames without thoughtful consideration.
In the expert-evaluation phase, the participants' task was to review 36 problem frames each, randomly sampled from the distinct reframing approaches and problems (4 approaches $\times$ 3 problems $\times$ 3 frames).

\begin{table*}[t]
    \centering
    \caption{Examples of high- and low-quality problem frames created in the study}
    \renewcommand{\arraystretch}{1.2} 
    \begin{tabularx}{\textwidth}{>{\hsize=0.23\hsize}XXX}
    \multicolumn{3}{p{\textwidth}}{\textbf{Problem:} In an office, employees have established a tradition of having lunch together, which has strengthened their team bond and promoted social interactions. However, since the number of employees has increased, choosing a lunch spot that can satisfy everyone has become quite challenging. Now, employees face multiple conflicting preferences, such as food quality over price or distance to restaurants and vice versa. These considerations often lead to having an excessively long discussion or everyone having lunch separately, undermining their valued lunchtime culture.} \vspace{0.25cm}\\    
    \toprule
     & \textbf{High novelty} & \textbf{Low novelty} \\
    \midrule
    \textbf{High \newline usefulness} & ``If the problem of choosing a lunch spot that satisfies everyone is approached as if the problem of gamifying group decision-making, then the direction of solutions could be creating a `Lunch Choice Challenge' where employees earn points or rewards by participating in various lunch outings and voting on future locations.'' & ``If the problem of choosing a lunch spot that satisfies everyone is approached as if the problem of how to rotate days between workers for lunch spots, then the direction of solutions could be to make a schedule for workers lunchtimes.'' \vspace{1cm} \\    
    \textbf{Low \newline usefulness} & ``If the problem of employee growth making it impossible to find/create a new lunch location that satisfies everyone is approached as if an opportunity instead of a problem, then the direction of the solutions could be employees eating at their work stations and using Zoom to talk to each other.''  & ``If the problem of undermining the valuable lunchtime culture is approached as if the problem of not eating together as a team, then the direction of solutions could be to no longer have the same lunch as each other.'' \\
    \bottomrule
    \end{tabularx}
    \label{tab:example_frame}
\end{table*}

\paragraph{A template for filling in problem frames.}
We defined the template for reporting problem frames. When participants evaluate problem frames, they should focus on the alternative perspectives on problems each frame introduces.
Without providing a clear template, participants might write frames in varying lengths and styles, which could lead to similar perspectives being perceived as different or vice versa. Moreover, the variance in participants' writing skills could influence how effectively they express their perspectives in the frames.
To mitigate these issues, we adopted Dorst's problem frame format that includes the initial problem, alternative perspectives, and potential direction of solutions (i.e., \textbf{If the problem of} \{initial problem\} \textbf{is approached as if} \{alternative perspective\}, \textbf{then the direction of solutions could be} \{potential solutions\}) \cite{dorst:book:reframing}.

\subsection{Metrics}
The key metrics for testing our hypotheses were
\begin{itemize}
    \item the novelty and usefulness of the problem frames,
    \item perceived agency and ownership in reframing with LLMs,
    \item the designers' competence in problem-reframing, and
    \item the perceived helpfulness of LLMs.
\end{itemize}
In empirical research on creativity, idea quality is evaluated by asking creative practitioners to assess them \cite{james:2022:evaluating_creativity, sarkar:2011:assessing_design_creativity, long:2024:novelty_ai_workflow}. We too followed \rr{this expert-rating-based approach} and adopted questionnaires designed to assess the novelty and utility of ideas \cite{doshi:2024:llm_writing_bad}. \rr{Our questionnaires prompted experts to assess how novel and useful a perspective each problem frame introduces when compared to the original problem description.}
To measure participants' agency and ownership, in turn, we employed \rr{self-assessment} items used with human--AI collaboration systems (e.g., ``How much do you feel that you were reframing the problem?'' and ``How much do you feel that you owned the problem frames?'') \cite{chan:2022:positive_negative_agency}.

\rr{We judged designers' competence by their conceptual and procedural understanding of problem reframing. This is based on the evidence that designers' ability to reframe problems correlates positively with their understanding of problem reframing \cite{gao:pf_collaboration, dorst:book:reframing, silk:pf_cognitiveStyle, cross:2004:design_expertise}.}
For this purpose, we designed a quiz with five multiple-choice questions based on the key concept, activity, and outcome of problem reframing \cite{dorst:book:reframing, schon:book:reflective, crilly:2021:co-evolution} where each question had four response options, \rr{among which were 1--2 correct answers (see Appendix \ref{appendix:quiz})}.
Finally, for the LLM-based approaches, we measured the participant-perceived helpfulness of the LLMs by using the Creativity Support Index (CSI) \cite{csi}.
To understand participants' experience more deeply, we probed perceived effort via NASA-TLX \cite{hart:2006:nasa} and collected qualitative reflections on the challenges they faced during the study's problem-reframing via open-ended questions.

\begin{figure*}[t]
    \centering
    \includegraphics[width=\linewidth]{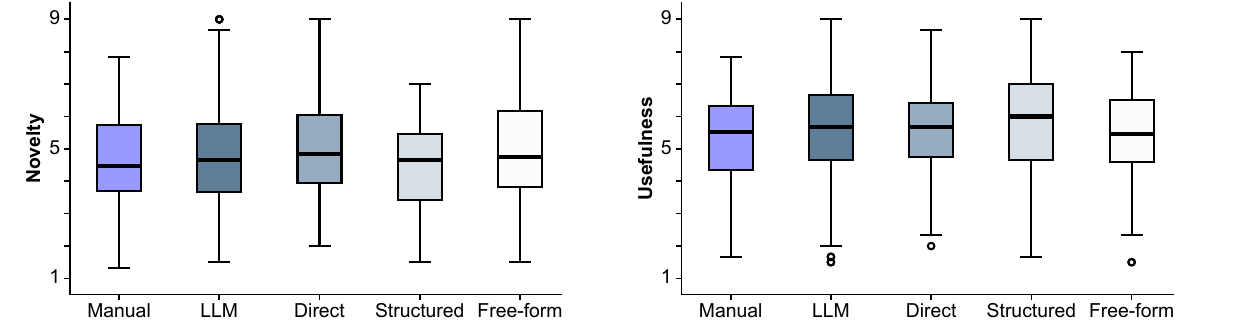}
    \caption{Using LLMs in problem reframing increased neither the novelty nor the usefulness of the problem frames, as evaluated by expert designers.}
    \Description{Two boxplots showing the result of analyzing the expert-evaluated novelty and usefulness of problem frames.}
    \label{fig:h1_n_u}
\end{figure*}

\begin{figure*}[t]
    \centering
    \includegraphics[width=\linewidth]{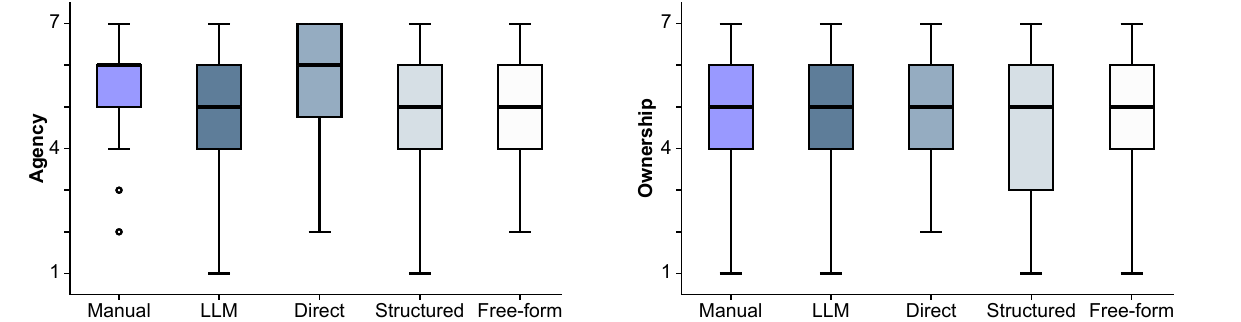}
    \caption{Participants using LLMs in problem reframing did not perceive less agency in the process or decreased ownership of problem frames.}
    \Description{Two boxplots showing the result of analyzing participants' perceived agency and ownership in problem reframing.}
    \label{fig:h2_a_o}
\end{figure*}

\subsection{Procedure}
Conducting the study on crowdsourcing platforms, we strove to collect high-quality responses by using their screening systems to recruit individuals with a background in design and a history of trustworthy behavior\footnote{~This is discussed at \url{https://www.prolific.com/resources/find-filter-favourite-how-to-select-participants-for-ai-tasks}.}.
In addition, as an incentive to generate high-quality problem frames, we offered a bonus (of 5 euros) to the 10\% who generated the most novel and useful frames.
Our reframing phase comprised three parts: 1) an initial survey, 2) problem reframing, and 3) a final survey.
The pre-survey asked about participants' demographics, experience with AI systems, and level of understanding of problem reframing.
Completing the survey sent the participants automatically to our interface, where they were randomly assigned a reframing approach and problem.

Across all approaches, our interface gave a uniform description and example of problem-reframing, also setting the same goal and evaluation criteria for all participants.
For the \textsc{direct} and \textsc{structured} approach, participants were given a tutorial in how to use our system.
Participants assigned the \textsc{free-form} approach were instructed to \rr{use the default version of ChatGPT with their account (the model was the same for all users: GPT-4o). We confirmed its use by having them submit their conversation log.}
Each participant submitted the final problem frames through our system and completed the post-survey.
Prior to the evaluation phase, we assessed the participants' expertise via the quiz from the pre-survey. 
Those participants we considered to be experts were invited to phase 2. The 15 individuals who responded evaluated the quality of problem frames generated by other participants.

\section{Results}
On average, the participants created two problem frames \rr{in 23.25 minutes ($SD$ = 14.73) in the \textsc{manual}, 21.82 minutes ($SD$ = 11.95) in the \textsc{direct}, 28.35 minutes ($SD$ = 18.00) in the \textsc{structured}, and 20.78 minutes ($SD$ = 9.34) in the \textsc{free-form} condition.}
\rr{In the \textsc{structured} approach, all participants except one entirely regenerated all content at least once. They continued to regenerate individual content, mainly the apparent problem (\emph{n}~=~39) and stakeholders' perspective (\emph{n}~=~25). Similarly, participants in the \textsc{direct} approach repeatedly regenerated problem frames, apart from nine who generated frames only once.}
We analyzed the 280 problem frames that participants selected as the best ones from their reframing.
The example problem frames and their qualities are shown in Table \ref{tab:example_frame}.

To test each hypothesis, we conducted statistical analysis on two levels. Firstly, to understand the influence of using LLMs relative to not using them, we took the LLM-based approaches as a single group (\textsc{llm-all}) for comparison to the \textsc{manual} approach. For this, we performed Mann--Whitney $U$ testing, considering the imbalance in sample size between the groups.
Secondly, to understand the influence of different ways of using LLMs, we made comparisons among the \textsc{manual}, \textsc{direct}, \textsc{structured}, and \textsc{free-form} approaches. Here, we applied the Kruskal–Wallis test, a non-parametric test designed to compare more than two independent groups. We performed Dunn's test with Bonferroni correction as \emph{post-hoc} testing. In all cases, we considered $p$~<~0.05 to reflect a statistically significant observation.

\subsection{Power Analysis}
We used the G*Power application to compute the minimum sample size required to detect medium effect size with 95\% power. The first and second analyses require sample sizes of 220 and 280 participants, respectively. Accordingly, we confirm that we had enough participants (at \emph{n}~=~280) for both analyses.

\begin{figure*}[t]
    \centering
    \includegraphics[width=\linewidth]{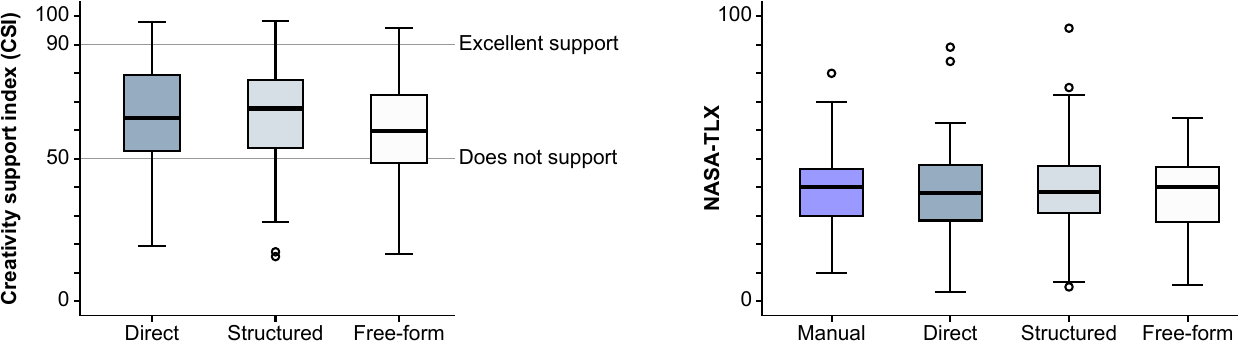}
    \caption{The \textsc{structured} approach did not increase participants' perceived helpfulness of LLMs in problem reframing (left). All LLM-based approaches were within the range indicating good support in creative activities \cite{csi}. Participants' perceived effort in the process also did not statistically differ between the approaches (right).}
    \Description{Two boxplots showing the result of analyzing creativity support index and NASA-TLX.}
    \label{fig:h3_c_n}
\end{figure*}

\begin{figure*}[t]
    \centering
    \includegraphics[width=\linewidth]{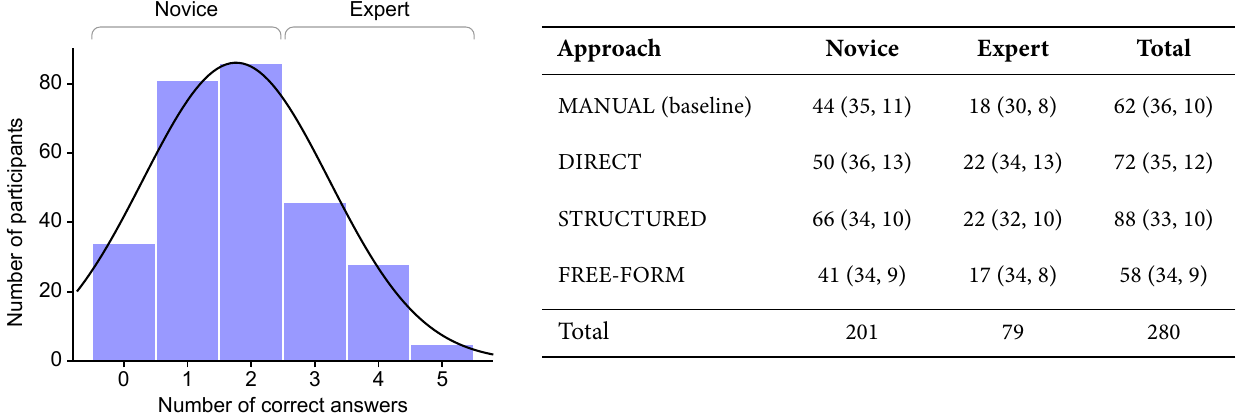}
    \caption{The distribution plot on the left shows how many questions about problem reframing the participants answered correctly \rr{(see Appendix \ref{appendix:quiz} for correct answers)}. The associated number of participants decreases dramatically after two correct answers. Accordingly, We categorized people who gave three or more correct answers as experts and the rest as novices. The table on the right gives a breakdown of the experts and novices by approach. Their age $Mean$ and $SD$ are shown in the bracket.}
    \Description{A distribution plot and a table showing how we categorized participants as experts or novices.}
    \label{fig:expertise}
\end{figure*}

\subsection{Using LLMs Did Not Help Participants Generate More Novel and Useful Problem Frames [H1]}
\label{ss:result_h1}

We compared the novelty and usefulness of problem frames. As Figure \ref{fig:h1_n_u} shows, there were no statistically significant differences for either, between the \textsc{manual} and the \textsc{llm-all} group but also among all the individual approaches (all $p$~>~0.05).
This indicates that reframing problems with LLMs did not help designers generate more novel or useful problem frames in comparison to working alone.
Our post survey suggests points to possible reasons for this. 
Of the 218 participants in the \textsc{llm-all} group, 93 reported challenges in building on the ideas generated by LLMs. The most commonly cited reasons were that the LLM-generated ideas were not novel (\emph{n}~=~31), repetitive (\emph{n}~=~14), and not useful in solving problems (\emph{n}~=~12). 
For example, P275 commented, \textit{``I just have this nagging feeling that it just can't really grasp the `why' of it all. It doesn't understand `why' the lunchtime culture is important ... It can come up with very practical outcomes, but it doesn't grasp human behavior well; for example, the need for spontaneity and for keeping the lunch casual is not something it could or would have picked up on its own.''} 
Accordingly, we reject \textbf{H1} and conclude that using LLMs does not help designers create better problem frames.

\begin{figure*}[t]
    \centering
    \includegraphics[width=0.95\linewidth]{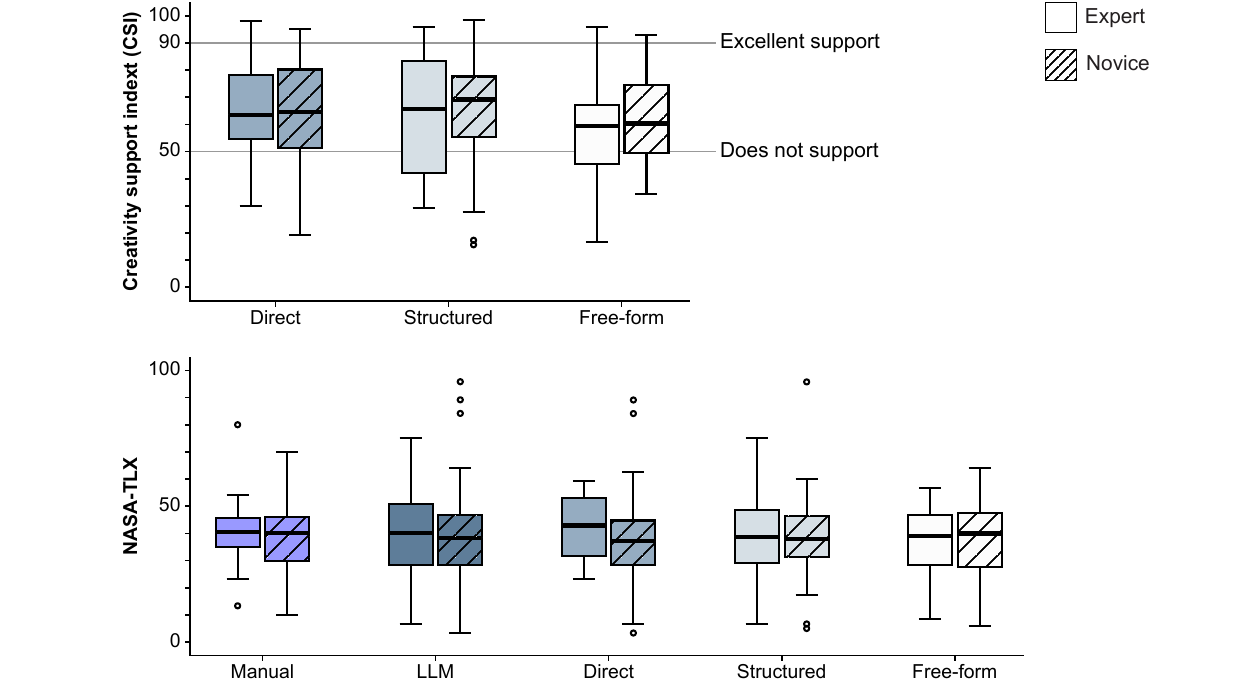}
    \caption{Neither perceived helpfulness of LLMs (top) nor effort in problem reframing (bottom) differed statistically significantly between designers with high (expert) and low understanding of problem reframing (novice).}
    \Description{Two boxplots showing the result of analyzing the creativity support index and NASA-TLX between experts and novices.}
    \label{fig:h4_competence_2}
\end{figure*}

\subsection{Using LLMs Did Not Decrease Agency and Ownership [H2]}
\label{ss:result_h2}

We compared participants' perceived agency in the reframing process and their ownership of the problem frames, conducting an analysis similar to that described above. Figure \ref{fig:h2_a_o} shows the results. Here too, there were also no statistically significant differences, whether between the \textsc{manual} and \textsc{llm-all} groups or across all approaches (all $p$ > 0.05). This indicates that participants retained their sense of agency and of ownership despite the use of LLMs and that the manner in which they interacted with LLMs did not significantly affect these perceptions.

Our post-survey also revealed that participants showed relatively little concern about agency and ownership. From among the 93 participants who reported challenges in reframing with LLMs, only 12 expressed a general dislike for AI when thinking about loss of agency/ownership.
\rr{While this number is small, their concerns do spotlight the fundamental issue of replacing creative inputs with LLMs. For example, P74, exposed to the \textsc{structured} approach, commented, \textit{``AI made it (problem reframing) easier. I didn't have to think very hard since the AI did everything ... It made me feel like there was no point in my trying to solve the problem at all.''}}
%
%
In summary, we rejected \textbf{H2} and concluded that using LLMs does not decrease designers' felt agency and ownership in problem reframing.
\rr{Yet, we note that further investigations are needed to uncover the potential loss of human creativity, which designers might not recognize amid the convenience of using LLMs.}

\subsection{Structured Use of LLMs Did Not Increase Perceived Helpfulness of LLMs [H3]}
\label{ss:result_h3}

Next, we looked at participants' perceptions of the helpfulness of LLMs (Figure \ref{fig:h3_c_n}).
We collected CSI scores for the LLM-based approaches only, since this metric is designed to measure the perceived helpfulness of a given system for creative activities \cite{csi} and hence is not applicable to the \textsc{manual} approach. 
The CSI score for the \textsc{direct}, \textsc{structured}, and \textsc{free-form} approach was 64.28 ($SD$~=~18.90), 64.85 ($SD$~=~20.00), and 61.01 ($SD$~=~17.19), respectively.
The index's developers \cite{csi} define scores between 50 and 90 as implying good support in creative activities.
Accordingly, our results show that participants perceived all LLM-assisted approaches as helpful. 
However, there were no statistically significant differences between approaches.
The NASA-TLX analysis echoes this finding. There was no statistically significant difference in how much effort participants felt, over all approaches.

\rr{While these findings suggest that the differences in ways of using LLMs did not influence how much assistance was perceived from LLMs, participants' comments suggest that each approach might be unique in the primary type of assistance sensed. Whereas the benefit reported most for the \textsc{direct} and the \textsc{free-form} condition (with \emph{n}~=~26 and \emph{n}~=~13, respectively) lay in helping diversify perspectives, participants' report on the \textsc{structured} approach emphasized helping examine the problem (\emph{n}~=~27). Representing this well, P138 commented that the \textit{``prompts and examples offered by the AI helped me think more critically about different aspects of the problem and explore solutions that I might not have considered otherwise.''}  We discuss differences of this sort later in the paper.}
In summary, we rejected \textbf{H3}. The \textsc{structured} approach to using LLMs did not increase designers' sense that LLMs are helpful in comparison to the other approaches.

\begin{figure*}[t]
    \centering
    \includegraphics[width=0.95\linewidth]{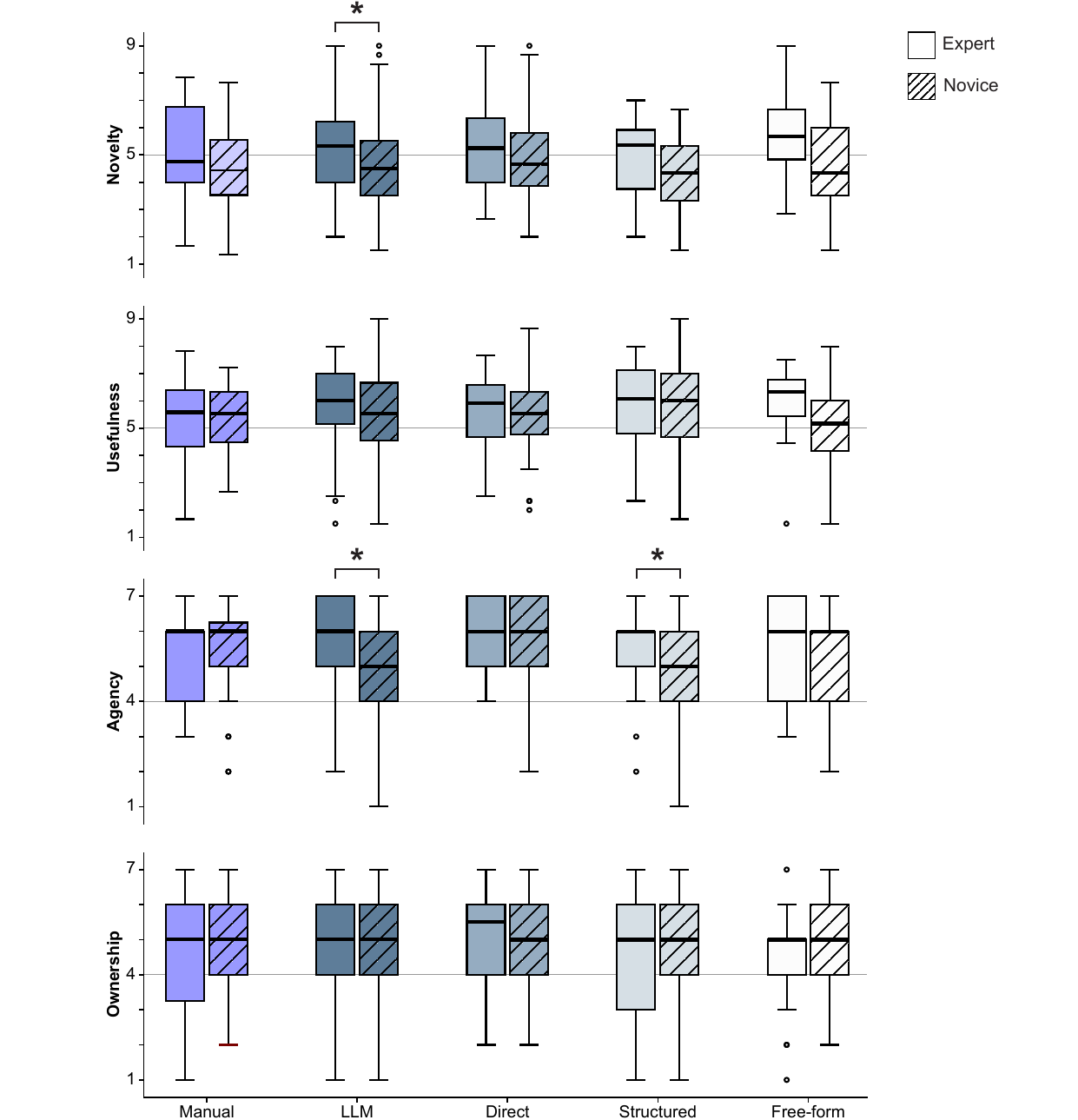}
    \caption{LLM use exhibited benefits for those designers with a higher understanding of problem reframing. Experts could generate more novel problem frames and perceive greater agency than novices did.}
    \Description{Four boxplots showing the result of analyzing novelty, usefulness, agency, and ownership between experts and novices.}
    \label{fig:h4_competence_1}
\end{figure*}

\subsection{Using LLMs Increased the Competence Gap Between experts and novices [H4]}
\label{ss:result_h4}

Lastly, we carried out all of the aforementioned analyses also between participants with different levels of expertise in problem reframing.
For this, we based our categorization of participants as experts or novices on their number of correct answers in the quiz. As the plot on the left in Figure \ref{fig:expertise} shows, there is a distinct separation at the point of having answered at least three questions correctly. Therefore, we classified participants who answered three or more questions correctly as experts and the rest as novices. A chi-squared test revealed that the distribution of experts and novices across the reframing approaches was not statistically significant, $\chi^{2}$(3, $N$ = 280) = 0.70, $p$ = 0.87, indicating that assignment in this study was balanced well for participants' variation in expertise. A breakdown of the participants by condition is provided at the right in Figure \ref{fig:expertise}.


Our comparison between experts and novices is characterized in figures \ref{fig:h4_competence_2} and \ref{fig:h4_competence_1}. As with the results reported above, there were no statistically significant differences in perceptions of helpfulness or of effort: for experts vs. novices, all $p$ > 0.05. Both groups' CSI scores across the LLM-based approaches lay within the moderate-helpfulness range.
This suggests that designers with different levels of expertise in problem reframing perceived similar levels of helpfulness from LLMs.

Next, we compared the novelty and usefulness of the problem frames between the experts and novices. 
We identified statistically significant differences in the novelty of the LLM-conditioned frames ($Z$ = 2.70, $p$ = 0.01): the experts generated 12.54\% more novel frames than the novices.
While the novelty gap was largest in the \textsc{free-form} approach (at 20.84\%), that size difference was not statistically significant ($Z$ = 2.31, $p$ = 0.06). As for the usefulness of frames, no statistically significant differences emerged across the approaches (all $p$ > 0.05).
This suggests that using LLMs might expand the novelty gap among designers: those who are stronger at problem reframing could end up creating frames that demonstrate more novelty than novices' do. However, which facets of LLM-based approaches contribute to the gap, and how, requires further investigation.

Lastly, we compared the perceived agency and ownership between the experts and novices. Analysis showed that experts sensed greater agency from using LLMs than novices did, in statistical terms ($Z$ = 2.05, $p$ = 0.04). We found that the gap was especially significant for the \textsc{structured} condition ($Z$ = 2.29, $p$ = 0.02), wherein the experts felt 15.69\% more agency than the novices. 
In contrast, no statistically significant differences were visible in how the experts vs. novices perceived the ownership of their frames across the approaches (all $p$ > 0.05).
Accordingly, \textbf{H4} was supported, and we conclude that using LLMs can confer greater benefits for those designers with higher expertise in problem reframing as compared to those with lower expertise. 
Our findings add a new layer to the picture from work showing that LLM assistance can close the gap between people with differing expertise/creativity levels \cite{noy:2023:llm_writing_good, doshi:2024:llm_writing_bad} by helping less design-competent people arrive at high-quality outcomes as more experienced individuals do. We unpack this finding in the discussion section.


\section{Discussion}


Advancements in generative AI have sparked both excitement and skepticism in the HCI and design communities \cite{si:2024:llm_novel_research, li:2024:ux_perception_ai}.
While there is justified excitement stemming from LLMs' promise to augment human creativity, evidence as to whether LLMs can yield notable improvements remains mixed.
With this paper, we have offered empirical evidence that using LLMs does not confer advantages in designers' problem reframing over not using them.
Our main finding is that using them, in several distinct approaches, had no impact on the novelty or usefulness of the problem frames.
Instead, we found that using LLMs increased the competence gap between designers with high and low understanding of problem reframing.
Proceeding from our findings, we conclude that LLMs do not offer a clear advantage in the reframing. We discuss the main reasons for this lack of impact \rr{and routes for further verifying our findings}.

%
%



\subsection{Why Using LLMs Did Not Benefit Designers in Problem Reframing?}
We review our three hypotheses with previous studies to understand the lack of LLMs' impact in our study. 
The hypotheses and results are summarized side by side in Table \ref{tab:result_overview}.

\begin{table*}[t]
    \centering
    \caption{The study's hypotheses and results in summary}
    \Description{A table showing the summary of our hypotheses and study results.}
    \renewcommand{\arraystretch}{1.2} 
    \begin{tabularx}{\textwidth}{XcX}
    \toprule
    \textbf{Hypotheses} & \textbf{Conf.} & \textbf{Study result} \\
    \midrule
    \textbf{H1}: Using an LLM during problem reframing increases the novelty and usefulness of the outputs. & \xmark & No increase in novelty or usefulness from the manual to LLM-based approaches. \\
    \textbf{H2}: Using an LLM during problem reframing decreases felt agency and ownership. & \xmark & No decrease in agency and ownership between the manual and LLM-based approaches. \\
    \textbf{H3}: Guiding designers to use an LLM in a structured manner increases their perceived helpfulness of LLMs. & \xmark & No differences in CSI and NASA-TLX between the approaches. \\
    \textbf{H4}: Using an LLM benefits designers with higher level of expertise in problem reframing more. & \cmark & More competent designers generate more novel frames and perceive higher agency. \\
    \bottomrule
    \end{tabularx}
    \label{tab:result_overview}
\end{table*}

Firstly, considering \textbf{H1}, we can assume that the quality of the ideas generated by LLMs might have been why they did not help designers create more novel/useful frames.
Numerous studies have explored LLMs' potential to support creative tasks by generating diverse ideas \cite{lin:2024:ai_speech_writing, chakrabarty:2024:llm_writers, yuan:2022:wordcraft_writing, he:2024:ai_groupIdeation, llm_creativity:dis:2022}. The ideal expressed is that LLM-generated material inspires people to generate more creative ideas that they might not have considered otherwise \cite{share:2024:brainwriting_llm, suh:2024:luminate, qin:2024:ai_character_writing}.
Yet contrasting evidence, pointing to LLM-generated ideas as not always original enough to spark improvement on the ideas \cite{doshi:2024:llm_writing_bad, anderson:2024:homogen_llm_ideation, koivisto:2023:human_outperform_ai}, is more consistent with our observations.
Also, answers to our post-experiment questions from 34\% of respondents identify the challenges as arising from the LLM-generated frames not being novel bases for them to build upon.
Therefore, at least in the context of reframing the design problems posed, LLMs might not be competent at generating perspectives that can inspire designers to any significant degree.


Secondly, the designers' retention of their perceived agency and ownership, counter to \textbf{H2}, might well be tied to the nature of problem reframing. Problem reframing is inherently a designer-driven activity wherein users must reflect on their understanding of the problem at hand \cite{dorst:book:reframing, schon:book:reflective, silk:pf_cognitiveStyle}; therefore, participants might have perceived their use of LLM-generated content as an extension of their own reflective process \cite{wen:2022:sense_of_agency}. This reflective element might also explain why the designers maintained a sense of ownership over the resulting problem frames. They may have perceived the frames as products of their deeper understanding of the problems rather than contributions from LLMs.
While a few studies have shown that generating ideas with LLMs can decrease individuals' felt agency \cite{guo:2024:agency_llm, moruzzi:2022:AI_agency, lawton:2023:co-drawing_agency}, we believe that there is a fundamental difference between idea generation and sophisticated reflective practice -- such as that embodied in problem reframing.

Thirdly, with regard to \textbf{H3}, we could connect the similarity in the perceived helpfulness of LLMs across approaches to the ``irony of automation.''
This notion suggests that systems designed to automate users' tasks might, rather than reduce their effort, merely redirect them to another task, such as reviewing systems' output \cite{baxter:ecce:2012irony}.
Our initial assumption was that participants would perceive the \textsc{structured} approach to be more helpful than the other approaches since it provides intermediate content in each problem reframing step \cite{dorst:book:reframing}. Indeed, previous studies have shown that such structured ways of using LLMs are perceived to be helpful in creative activities \cite{suh:2024:luminate, xu:2024:llm_japmlate_why}.
In contrast, we found that participants sensed similar levels of helpfulness from all the approaches and comparable problem-reframing efforts, even relative to not using LLMs.
Therefore, while the type of effort invested by participants might hinge on the approach (e.g., reviewing LLM-generated content in the \textsc{structured} approach vs. refining LLM-generated frames in the \textsc{direct} one), we assume that the total amounts of effort may have been similar.

\subsection{Why Using LLMs Increased the Competence Gap in Problem Reframing?}

An optimistic view of generative AI in diverse fields has been that LLMs can help less experienced people achieve high-quality outcomes similar to more experienced individuals' \cite{bryn:2023:ai_help_novice}, thus narrowing the gap between experts and novices.
Prior studies likewise suggest that using LLMs helps people with lower creativity levels \cite{doshi:2024:llm_writing_bad} or novice writers \cite{noy:2023:llm_writing_good} significantly improve their outcomes, while more competent individuals gain little improvement from using LLMs.
We could assume, then, that LLM-generated ideas are not as high-quality as what experts can generate \cite{long:2024:novelty_ai_workflow, chakrabarty:2024:llm_false_promise_art}; hence, they may not offer substantial benefit to experts.

In contrast, we found that using LLMs in problem reframing can increase the gap between experts and novices.
Our results related to \textbf{H4} suggest that using it exerted an influence whereby designers with more advanced understanding of problem reframing generated more novel frames and sensed greater agency than less competent designers did.
This gulf in frame novelty is consistent with our initial hypothesis -- experts, with their deep understanding of the reframing process, are better equipped to evaluate and refine LLM-generated ideas.
Considering prior evidence that designers with less competence in problem-reframing tend to tackle the problem as given \cite{silk:pf_cognitiveStyle, cross:2004:design_expertise}, we can assume that the novices may have struggled to move beyond the initial suggestions from LLMs and were more likely to accept them without further exploration.

One especially noteworthy observation from our study is that guiding designers to use LLMs in a \textsc{structured} manner led more competent ones to perceive greater agency than felt by less competent individuals.
In previous studies, enabling structured ways to use LLMs has been preferred \cite{suh:2024:luminate, xu:2024:llm_japmlate_why, shin:2024:persona}, for their provision of clear steps and guidelines that support users' better incorporation of AI-generated suggestions into their workflows.
Accordingly, we may conclude that the \textsc{structured} approach -- built on Dorst's nine-step process \cite{dorst:book:reframing} -- is likely to gibe with more competent designers' reframing workflow and, thereby, leave them experiencing more control over the process. In contrast, designers who are less competent at problem reframing might not be as familiar with this structured process so might be led to feel that they are following set guidelines rather than directing the workflow themselves.





\subsection{Implications for Practice}

We advise against simplistically relying on LLMs as standalone creativity-support tools in problem reframing.
Across the various metrics applied in this study, there was not a single setting wherein using LLMs yielded significant benefits for designers relative to not using them.
Notably, while the analysis of CSI scores showed that participants perceived LLMs as helpful, this perception did not translate into generating more novel or useful problem frames.
Especially for inexperienced designers, thoughtless use of LLMs may compromise the novelty of the frames and impair felt agency. 

Nonetheless, exploiting LLMs as auxiliary tools may be beneficial.
The most commonly reported issue with this study's LLM-generated frames was that they were too low-quality to serve as foundations.
That limitation could act as an advantage, though.
Because LLMs generate many low-quality frames, designers may be able to learn from that output when trying to discern what renders a problem hard.
This way of using LLMs dovetails with Dorst's step for identifying previous, ineffective solutions in aims of understanding what the problems are \cite{dorst:book:reframing}.

\subsection{Limitations and Future Work}
%


\rr{
Our study had four main limitations.
Firstly, the tests used only GPT-4o. While it is a state-of-the-art LLM building on GPT-4 \cite{openai:2023:gpt4}, it might not be the best to reframe design problems with.
Secondly, there could be other ways of using LLMs, ones not tested in this study. For example, the \textsc{free-form} and \textsc{structured} approaches could be combined such that LLMs converse with designers to guide them through a structured process.
Also, our study design might have influenced how participants approached problem reframing.
By instructing them to submit the best-quality frames, we might have been seen as prioritizing the elimination of low-quality frames rather than broadening of the frames' range.
Furthermore, evaluation of problem frames might differ in several respects between those who deeply understand problem reframing and those whose competence is centered on evaluation.
The final possible confounding factor is related to inherent limitations of the crowdsourcing approach. Irrespective of the sample sizes they afford, crowdsourcing platforms might not represent rich designer-population resources. Compared to in-lab studies, crowdsourcing provides a less detail-oriented view of participants' behaviors.
}

\rr{
In light of these limitations, we propose the following directions for research that could further test the generalizability of our findings and, potentially, uncover new ways in which LLMs might enhance problem reframing.}
\rr{
In the first, future work could look into the diverse competencies of designers.
We found that LLMs are more beneficial for designers who possess a stronger understanding of problem reframing.
The differential impact of LLMs could be further explored in relation to other competencies too, such as designers' conceptual-level understanding in practice and their facility with outcomes.
For example, research could yield a more nuanced view of LLMs that clarifies what distinguishes designers who know what problem reframing is but struggle to reframe problems from those who can derive high-quality frames without a solid understanding of problem reframing.
For this, we recommend finer-granularity work on the elements of design expertise required in problem reframing.
}

\rr{
Second, future work should explore LLMs' influence on alternative qualities of problem frames.
We found that the LLM-based approaches do not offer the advantage of generating more novel and useful problem frames;
however, a designer might prioritize other aspects of frames, such as highly evocative metaphors for a given problem or power to persuade multiple stakeholders \cite{dorst:book:reframing}.
Work focusing on how designers can obtain such qualities with LLMs could pinpoint alternative ways of using them, at which LLMs might be more adept than generating original concepts.
We believe research could explore this avenue by varying the goal set for the reframing over multiple studies. 
}

\rr{
Future work also should investigate how exactly designers utilize LLM-generated frames.
We have taken the first steps toward understanding whether using LLMs benefits designers or not, but we need 
further observations of what designers do with LLM-generated content before we can fully understand the mechanisms behind what we saw.
Potentially, further work could reveal how designers adapt frames generated from LLMs, whether they prematurely filter out useful perspectives developed with LLMs, etc.
We recommend studying these factors in laboratory settings, where designers' step-by-step process of using LLMs can be observed in fine detail.
}

\rr{
Finally, we recommend studying the use of LLMs in real-world practice.
Our study stressed the ideation phase of problem reframing, but design practice also entails engaging in several other tasks related to reframing, such as gathering details about stakeholders and negotiating with teams \cite{lee:2020:pf_convince}.
Often, said tasks are performed in multiple events, throughout a design project \cite{dorst:book:reframing}.
Investing in uses of LLM over the entire lifecycle of problem-reframing could expand our observations' richness and utility. For example, designers might be able to apply LLMs to articulate the value of the new-framing problems for stakeholders or to reach a realistic level of framing in practice.
To enable observations on this front, scholars could conduct longitudinal studies able to identify the effects of using LLMs, for a vast range of reframing scenarios.
}

\section{Conclusions}
In this paper, we explored the influence of using LLMs in problem reframing. 
Building on the conflicting evidence on generative AI's assistance in creative activities, we formulated four hypotheses previously untested in the context of problem reframing.
Our hypotheses focused on how three distinct LLM-based approaches would affect designers with varying levels of competence in problem reframing. Specifically, we examined how the use of LLMs influenced the novelty and usefulness of the problem frames, as well as designers' perceived agency and ownership.
With 280 designers, we rigorously compared the LLM-based approaches to the baseline condition, where designers reframed problems without support from LLMs.
We found that using LLMs does not benefit designers in generating high-quality problem frames. 
The resulting frames were of similar qualities compared to those not using LLMs.
Rather, we observed the potential risk of using LLMs, increasing the gap between the designers with more and less competence in problem reframing.
Based on our findings and the literature behind our hypotheses, we discussed the main reasons for the lack of effect from using LLMs.
In sum, our study enriches empirical understanding of utilizing LLMs in designerly activities.


\begin{acks}
This research was supported by the Research Council of Finland project Subjective Functions (357578), Finnish Center for Artificial Intelligence (328400, 345604, 341763), European Research Council Advanced Grant (101141916), and the Department of Art and Media and the Department of Information and Communications Engineering at Aalto University.
\end{acks}

\bibliographystyle{ACM-Reference-Format}
\bibliography{coDesignAI}


\begin{thebibliography}{87}


\ifx \showCODEN    \undefined \def \showCODEN     #1{\unskip}     \fi
\ifx \showDOI      \undefined \def \showDOI       #1{#1}\fi
\ifx \showISBNx    \undefined \def \showISBNx     #1{\unskip}     \fi
\ifx \showISBNxiii \undefined \def \showISBNxiii  #1{\unskip}     \fi
\ifx \showISSN     \undefined \def \showISSN      #1{\unskip}     \fi
\ifx \showLCCN     \undefined \def \showLCCN      #1{\unskip}     \fi
\ifx \shownote     \undefined \def \shownote      #1{#1}          \fi
\ifx \showarticletitle \undefined \def \showarticletitle #1{#1}   \fi
\ifx \showURL      \undefined \def \showURL       {\relax}        \fi
\providecommand\bibfield[2]{#2}
\providecommand\bibinfo[2]{#2}
\providecommand\natexlab[1]{#1}
\providecommand\showeprint[2][]{arXiv:#2}

\bibitem[zhe(2018)]%
        {zheng:2018:ownership_bias}
 \bibinfo{year}{2018}\natexlab{}.
\newblock \bibinfo{booktitle}{\emph{{Uncovering Ownership Bias: The Influence of Idea Goodness and Creativity on Design Professionals’ Concept Selection Practices}}}. \bibinfo{series}{International Design Engineering Technical Conferences and Computers and Information in Engineering Conference}, Vol.~\bibinfo{volume}{Volume 7: 30th International Conference on Design Theory and Methodology}.
\newblock
\urldef\tempurl%
\url{https://doi.org/10.1115/DETC2018-85964}
\showDOI{\tempurl}
\showeprint{https://asmedigitalcollection.asme.org/IDETC-CIE/proceedings-pdf/IDETC-CIE2018/51845/V007T06A016/2477027/v007t06a016-detc2018-85964.pdf}


\bibitem[Al'tshuller(1999)]%
        {altshuller:1999:triz}
\bibfield{author}{\bibinfo{person}{Genrikh~Saulovich Al'tshuller}.} \bibinfo{year}{1999}\natexlab{}.
\newblock \bibinfo{booktitle}{\emph{The innovation algorithm: TRIZ, systematic innovation and technical creativity}}.
\newblock \bibinfo{publisher}{Technical innovation center, Inc.}
\newblock


\bibitem[Amabile(1983)]%
        {amabile:1983:creativity_motivation}
\bibfield{author}{\bibinfo{person}{Teresa~M Amabile}.} \bibinfo{year}{1983}\natexlab{}.
\newblock \showarticletitle{The social psychology of creativity: A componential conceptualization.}
\newblock \bibinfo{journal}{\emph{Journal of personality and social psychology}} \bibinfo{volume}{45}, \bibinfo{number}{2} (\bibinfo{year}{1983}), \bibinfo{pages}{357}.
\newblock


\bibitem[Anderson et~al\mbox{.}(2024)]%
        {anderson:2024:homogen_llm_ideation}
\bibfield{author}{\bibinfo{person}{Barrett~R Anderson}, \bibinfo{person}{Jash~Hemant Shah}, {and} \bibinfo{person}{Max Kreminski}.} \bibinfo{year}{2024}\natexlab{}.
\newblock \showarticletitle{Homogenization Effects of Large Language Models on Human Creative Ideation}. In \bibinfo{booktitle}{\emph{Proceedings of the 16th Conference on Creativity \& Cognition}} (Chicago, IL, USA) \emph{(\bibinfo{series}{C\&C '24})}. \bibinfo{publisher}{Association for Computing Machinery}, \bibinfo{address}{New York, NY, USA}, \bibinfo{pages}{413–425}.
\newblock
\showISBNx{9798400704857}
\urldef\tempurl%
\url{https://doi.org/10.1145/3635636.3656204}
\showDOI{\tempurl}


\bibitem[Ashkinaze et~al\mbox{.}(2024)]%
        {ashkinaze:2024:human_ai_evolve_idea}
\bibfield{author}{\bibinfo{person}{Joshua Ashkinaze}, \bibinfo{person}{Julia Mendelsohn}, \bibinfo{person}{Li Qiwei}, \bibinfo{person}{Ceren Budak}, {and} \bibinfo{person}{Eric Gilbert}.} \bibinfo{year}{2024}\natexlab{}.
\newblock \bibinfo{title}{How AI Ideas Affect the Creativity, Diversity, and Evolution of Human Ideas: Evidence From a Large, Dynamic Experiment}.
\newblock
\newblock
\showeprint[arxiv]{2401.13481}
\urldef\tempurl%
\url{https://arxiv.org/abs/2401.13481}
\showURL{%
\tempurl}


\bibitem[Bardwell(1991)]%
        {bardwell:1991:pf_process}
\bibfield{author}{\bibinfo{person}{Lisa~V Bardwell}.} \bibinfo{year}{1991}\natexlab{}.
\newblock \showarticletitle{Problem-framing: A perspective on environmental problem-solving}.
\newblock \bibinfo{journal}{\emph{Environmental Management}}  \bibinfo{volume}{15} (\bibinfo{year}{1991}), \bibinfo{pages}{603--612}.
\newblock


\bibitem[Baxter et~al\mbox{.}(2012)]%
        {baxter:ecce:2012irony}
\bibfield{author}{\bibinfo{person}{Gordon Baxter}, \bibinfo{person}{John Rooksby}, \bibinfo{person}{Yuanzhi Wang}, {and} \bibinfo{person}{Ali Khajeh-Hosseini}.} \bibinfo{year}{2012}\natexlab{}.
\newblock \showarticletitle{The Ironies of Automation: Still Going Strong at 30?}. In \bibinfo{booktitle}{\emph{Proceedings of the 30th European Conference on Cognitive Ergonomics}} (Edinburgh, United Kingdom) \emph{(\bibinfo{series}{ECCE '12})}. \bibinfo{publisher}{Association for Computing Machinery}, \bibinfo{address}{New York, NY, USA}, \bibinfo{pages}{65–71}.
\newblock
\showISBNx{9781450317863}
\urldef\tempurl%
\url{https://doi.org/10.1145/2448136.2448149}
\showDOI{\tempurl}


\bibitem[Brynjolfsson et~al\mbox{.}(2023)]%
        {bryn:2023:ai_help_novice}
\bibfield{author}{\bibinfo{person}{Erik Brynjolfsson}, \bibinfo{person}{Danielle Li}, {and} \bibinfo{person}{Lindsey~R Raymond}.} \bibinfo{year}{2023}\natexlab{}.
\newblock \bibinfo{booktitle}{\emph{Generative AI at Work}}.
\newblock \bibinfo{type}{Working Paper} 31161. \bibinfo{institution}{National Bureau of Economic Research}.
\newblock
\urldef\tempurl%
\url{https://doi.org/10.3386/w31161}
\showDOI{\tempurl}


\bibitem[Chakrabarty et~al\mbox{.}(2024a)]%
        {chakrabarty:2024:llm_false_promise_art}
\bibfield{author}{\bibinfo{person}{Tuhin Chakrabarty}, \bibinfo{person}{Philippe Laban}, \bibinfo{person}{Divyansh Agarwal}, \bibinfo{person}{Smaranda Muresan}, {and} \bibinfo{person}{Chien-Sheng Wu}.} \bibinfo{year}{2024}\natexlab{a}.
\newblock \showarticletitle{Art or Artifice? Large Language Models and the False Promise of Creativity}. In \bibinfo{booktitle}{\emph{Proceedings of the CHI Conference on Human Factors in Computing Systems}} (Honolulu, HI, USA) \emph{(\bibinfo{series}{CHI '24})}. \bibinfo{publisher}{Association for Computing Machinery}, \bibinfo{address}{New York, NY, USA}, Article \bibinfo{articleno}{30}, \bibinfo{numpages}{34}~pages.
\newblock
\showISBNx{9798400703300}
\urldef\tempurl%
\url{https://doi.org/10.1145/3613904.3642731}
\showDOI{\tempurl}


\bibitem[Chakrabarty et~al\mbox{.}(2024b)]%
        {chakrabarty:2024:llm_writers}
\bibfield{author}{\bibinfo{person}{Tuhin Chakrabarty}, \bibinfo{person}{Vishakh Padmakumar}, \bibinfo{person}{Faeze Brahman}, {and} \bibinfo{person}{Smaranda Muresan}.} \bibinfo{year}{2024}\natexlab{b}.
\newblock \showarticletitle{Creativity Support in the Age of Large Language Models: An Empirical Study Involving Professional Writers}. In \bibinfo{booktitle}{\emph{Proceedings of the 16th Conference on Creativity \& Cognition}} (Chicago, IL, USA) \emph{(\bibinfo{series}{C\&C '24})}. \bibinfo{publisher}{Association for Computing Machinery}, \bibinfo{address}{New York, NY, USA}, \bibinfo{pages}{132–155}.
\newblock
\showISBNx{9798400704857}
\urldef\tempurl%
\url{https://doi.org/10.1145/3635636.3656201}
\showDOI{\tempurl}


\bibitem[Chan et~al\mbox{.}(2022)]%
        {chan:2022:positive_negative_agency}
\bibfield{author}{\bibinfo{person}{Liwei Chan}, \bibinfo{person}{Yi-Chi Liao}, \bibinfo{person}{George~B Mo}, \bibinfo{person}{John~J Dudley}, \bibinfo{person}{Chun-Lien Cheng}, \bibinfo{person}{Per~Ola Kristensson}, {and} \bibinfo{person}{Antti Oulasvirta}.} \bibinfo{year}{2022}\natexlab{}.
\newblock \showarticletitle{Investigating Positive and Negative Qualities of Human-in-the-Loop Optimization for Designing Interaction Techniques}. In \bibinfo{booktitle}{\emph{Proceedings of the 2022 CHI Conference on Human Factors in Computing Systems}} (New Orleans, LA, USA) \emph{(\bibinfo{series}{CHI '22})}. \bibinfo{publisher}{Association for Computing Machinery}, \bibinfo{address}{New York, NY, USA}, Article \bibinfo{articleno}{112}, \bibinfo{numpages}{14}~pages.
\newblock
\showISBNx{9781450391573}
\urldef\tempurl%
\url{https://doi.org/10.1145/3491102.3501850}
\showDOI{\tempurl}


\bibitem[Chen et~al\mbox{.}(2024)]%
        {chen:2024:trizgpt}
\bibfield{author}{\bibinfo{person}{Liuqing Chen}, \bibinfo{person}{Yaxuan Song}, \bibinfo{person}{Shixian Ding}, \bibinfo{person}{Lingyun Sun}, \bibinfo{person}{Peter Childs}, {and} \bibinfo{person}{Haoyu Zuo}.} \bibinfo{year}{2024}\natexlab{}.
\newblock \bibinfo{title}{TRIZ-GPT: An LLM-augmented method for problem-solving}.
\newblock
\newblock
\showeprint[arxiv]{2408.05897}
\urldef\tempurl%
\url{https://arxiv.org/abs/2408.05897}
\showURL{%
\tempurl}


\bibitem[Cherry and Latulipe(2014)]%
        {csi}
\bibfield{author}{\bibinfo{person}{Erin Cherry} {and} \bibinfo{person}{Celine Latulipe}.} \bibinfo{year}{2014}\natexlab{}.
\newblock \showarticletitle{Quantifying the Creativity Support of Digital Tools through the Creativity Support Index}.
\newblock \bibinfo{journal}{\emph{ACM Trans. Comput.-Hum. Interact.}} \bibinfo{volume}{21}, \bibinfo{number}{4}, Article \bibinfo{articleno}{21} (\bibinfo{date}{jun} \bibinfo{year}{2014}), \bibinfo{numpages}{25}~pages.
\newblock
\showISSN{1073-0516}
\urldef\tempurl%
\url{https://doi.org/10.1145/2617588}
\showDOI{\tempurl}


\bibitem[Cox et~al\mbox{.}(2023)]%
        {prompt/cox2023prompting}
\bibfield{author}{\bibinfo{person}{Samuel~Rhys Cox}, \bibinfo{person}{Ashraf Abdul}, {and} \bibinfo{person}{Wei~Tsang Ooi}.} \bibinfo{year}{2023}\natexlab{}.
\newblock \bibinfo{title}{Prompting a Large Language Model to Generate Diverse Motivational Messages: A Comparison with Human-Written Messages}.
\newblock
\newblock
\showeprint[arxiv]{2308.13479}


\bibitem[Crilly(2021)]%
        {crilly:2021:co-evolution}
\bibfield{author}{\bibinfo{person}{Nathan Crilly}.} \bibinfo{year}{2021}\natexlab{}.
\newblock \showarticletitle{The Evolution of “Co-evolution” (Part I): Problem Solving, Problem Finding, and Their Interaction in Design and Other Creative Practices}.
\newblock \bibinfo{journal}{\emph{She Ji: The Journal of Design, Economics, and Innovation}} \bibinfo{volume}{7}, \bibinfo{number}{3} (\bibinfo{year}{2021}), \bibinfo{pages}{309--332}.
\newblock
\showISSN{2405-8726}
\urldef\tempurl%
\url{https://doi.org/10.1016/j.sheji.2021.07.003}
\showDOI{\tempurl}


\bibitem[Cross(2004)]%
        {cross:2004:design_expertise}
\bibfield{author}{\bibinfo{person}{Nigel Cross}.} \bibinfo{year}{2004}\natexlab{}.
\newblock \showarticletitle{Expertise in design: an overview}.
\newblock \bibinfo{journal}{\emph{Design Studies}} \bibinfo{volume}{25}, \bibinfo{number}{5} (\bibinfo{year}{2004}), \bibinfo{pages}{427--441}.
\newblock
\showISSN{0142-694X}
\urldef\tempurl%
\url{https://doi.org/10.1016/j.destud.2004.06.002}
\showDOI{\tempurl}
\newblock
\shownote{Expertise in Design}.


\bibitem[de~Wynter et~al\mbox{.}(2023)]%
        {Wynter:arxiv:2023}
\bibfield{author}{\bibinfo{person}{Adrian de Wynter}, \bibinfo{person}{Xun Wang}, \bibinfo{person}{Alex Sokolov}, \bibinfo{person}{Qilong Gu}, {and} \bibinfo{person}{Si-Qing Chen}.} \bibinfo{year}{2023}\natexlab{}.
\newblock \showarticletitle{An evaluation on large language model outputs: Discourse and memorization}.
\newblock \bibinfo{journal}{\emph{Natural Language Processing Journal}}  \bibinfo{volume}{4} (\bibinfo{date}{sep} \bibinfo{year}{2023}), \bibinfo{pages}{100024}.
\newblock
\urldef\tempurl%
\url{https://doi.org/10.1016/j.nlp.2023.100024}
\showDOI{\tempurl}


\bibitem[Dorst(2018)]%
        {dorst:framingProcess}
\bibfield{author}{\bibinfo{person}{C Dorst}.} \bibinfo{year}{2018}\natexlab{}.
\newblock \showarticletitle{Mixing practices to create transdisciplinary innovation: A design-based approach}.
\newblock \bibinfo{journal}{\emph{Technology Innovation Management Review}} (\bibinfo{year}{2018}).
\newblock


\bibitem[Dorst(2015)]%
        {dorst:book:reframing}
\bibfield{author}{\bibinfo{person}{Kees Dorst}.} \bibinfo{year}{2015}\natexlab{}.
\newblock \bibinfo{booktitle}{\emph{Frame innovation: Create new thinking by design}}.
\newblock \bibinfo{publisher}{MIT press}.
\newblock


\bibitem[Dorst and Cross(2001)]%
        {dorst:co-evolve}
\bibfield{author}{\bibinfo{person}{Kees Dorst} {and} \bibinfo{person}{Nigel Cross}.} \bibinfo{year}{2001}\natexlab{}.
\newblock \showarticletitle{Creativity in the design process: co-evolution of problem–solution}.
\newblock \bibinfo{journal}{\emph{Design Studies}} \bibinfo{volume}{22}, \bibinfo{number}{5} (\bibinfo{year}{2001}), \bibinfo{pages}{425--437}.
\newblock
\showISSN{0142-694X}
\urldef\tempurl%
\url{https://doi.org/10.1016/S0142-694X(01)00009-6}
\showDOI{\tempurl}


\bibitem[Doshi and Hauser(2024)]%
        {doshi:2024:llm_writing_bad}
\bibfield{author}{\bibinfo{person}{Anil~R. Doshi} {and} \bibinfo{person}{Oliver~P. Hauser}.} \bibinfo{year}{2024}\natexlab{}.
\newblock \showarticletitle{Generative AI enhances individual creativity but reduces the collective diversity of novel content}.
\newblock \bibinfo{journal}{\emph{Science Advances}} \bibinfo{volume}{10}, \bibinfo{number}{28} (\bibinfo{year}{2024}), \bibinfo{pages}{eadn5290}.
\newblock
\urldef\tempurl%
\url{https://doi.org/10.1126/sciadv.adn5290}
\showDOI{\tempurl}
\showeprint{https://www.science.org/doi/pdf/10.1126/sciadv.adn5290}


\bibitem[Einarsson et~al\mbox{.}(2024)]%
        {einarsson:2024:pf_ChatGPT}
\bibfield{author}{\bibinfo{person}{Hafsteinn Einarsson}, \bibinfo{person}{Sigrún~Helga Lund}, {and} \bibinfo{person}{Anna~Helga Jónsdóttir}.} \bibinfo{year}{2024}\natexlab{}.
\newblock \showarticletitle{Application of ChatGPT for automated problem reframing across academic domains}.
\newblock \bibinfo{journal}{\emph{Computers and Education: Artificial Intelligence}}  \bibinfo{volume}{6} (\bibinfo{year}{2024}), \bibinfo{pages}{100194}.
\newblock
\showISSN{2666-920X}
\urldef\tempurl%
\url{https://doi.org/10.1016/j.caeai.2023.100194}
\showDOI{\tempurl}


\bibitem[Ericsson et~al\mbox{.}(2018)]%
        {ericsson:2018:experience_expertise}
\bibfield{author}{\bibinfo{person}{K~Anders Ericsson}, \bibinfo{person}{Robert~R Hoffman}, \bibinfo{person}{Aaron Kozbelt}, {and} \bibinfo{person}{A~Mark Williams}.} \bibinfo{year}{2018}\natexlab{}.
\newblock \bibinfo{booktitle}{\emph{The Cambridge handbook of expertise and expert performance}}.
\newblock \bibinfo{publisher}{Cambridge University Press}.
\newblock


\bibitem[Feng et~al\mbox{.}(2024)]%
        {feng:2024:canvil_llm_ux}
\bibfield{author}{\bibinfo{person}{K.~J.~Kevin Feng}, \bibinfo{person}{Q.~Vera Liao}, \bibinfo{person}{Ziang Xiao}, \bibinfo{person}{Jennifer~Wortman Vaughan}, \bibinfo{person}{Amy~X. Zhang}, {and} \bibinfo{person}{David~W. McDonald}.} \bibinfo{year}{2024}\natexlab{}.
\newblock \bibinfo{title}{Canvil: Designerly Adaptation for LLM-Powered User Experiences}.
\newblock
\newblock
\showeprint[arxiv]{2401.09051}
\urldef\tempurl%
\url{https://arxiv.org/abs/2401.09051}
\showURL{%
\tempurl}


\bibitem[Fogliato et~al\mbox{.}(2022)]%
        {fogliato:facct:2022}
\bibfield{author}{\bibinfo{person}{Riccardo Fogliato}, \bibinfo{person}{Shreya Chappidi}, \bibinfo{person}{Matthew Lungren}, \bibinfo{person}{Paul Fisher}, \bibinfo{person}{Diane Wilson}, \bibinfo{person}{Michael Fitzke}, \bibinfo{person}{Mark Parkinson}, \bibinfo{person}{Eric Horvitz}, \bibinfo{person}{Kori Inkpen}, {and} \bibinfo{person}{Besmira Nushi}.} \bibinfo{year}{2022}\natexlab{}.
\newblock \showarticletitle{Who Goes First? Influences of Human-AI Workflow on Decision Making in Clinical Imaging}. In \bibinfo{booktitle}{\emph{Proceedings of the 2022 ACM Conference on Fairness, Accountability, and Transparency}} (Seoul, Republic of Korea) \emph{(\bibinfo{series}{FAccT '22})}. \bibinfo{publisher}{Association for Computing Machinery}, \bibinfo{address}{New York, NY, USA}, \bibinfo{pages}{1362–1374}.
\newblock
\showISBNx{9781450393522}
\urldef\tempurl%
\url{https://doi.org/10.1145/3531146.3533193}
\showDOI{\tempurl}


\bibitem[Gao and Kvan(2004)]%
        {gao:pf_collaboration}
\bibfield{author}{\bibinfo{person}{Song Gao} {and} \bibinfo{person}{Thomas Kvan}.} \bibinfo{year}{2004}\natexlab{}.
\newblock \showarticletitle{An Analysis of Problem Framing in Multiple Settings}. In \bibinfo{booktitle}{\emph{Design Computing and Cognition '04}}, \bibfield{editor}{\bibinfo{person}{John~S. Gero}} (Ed.). \bibinfo{publisher}{Springer Netherlands}, \bibinfo{address}{Dordrecht}, \bibinfo{pages}{117--134}.
\newblock
\showISBNx{978-1-4020-2393-4}


\bibitem[Gero et~al\mbox{.}(2022)]%
        {llm_creativity:dis:2022}
\bibfield{author}{\bibinfo{person}{Katy~Ilonka Gero}, \bibinfo{person}{Vivian Liu}, {and} \bibinfo{person}{Lydia Chilton}.} \bibinfo{year}{2022}\natexlab{}.
\newblock \showarticletitle{Sparks: Inspiration for Science Writing using Language Models}. In \bibinfo{booktitle}{\emph{Proceedings of the 2022 ACM Designing Interactive Systems Conference}} \emph{(\bibinfo{series}{DIS '22})}. \bibinfo{publisher}{Association for Computing Machinery}, \bibinfo{address}{New York, NY, USA}, \bibinfo{pages}{1002–1019}.
\newblock
\showISBNx{9781450393584}
\urldef\tempurl%
\url{https://doi.org/10.1145/3532106.3533533}
\showDOI{\tempurl}


\bibitem[Gero et~al\mbox{.}(2024)]%
        {gero:2024:llm_sensemaking}
\bibfield{author}{\bibinfo{person}{Katy~Ilonka Gero}, \bibinfo{person}{Chelse Swoopes}, \bibinfo{person}{Ziwei Gu}, \bibinfo{person}{Jonathan~K. Kummerfeld}, {and} \bibinfo{person}{Elena~L. Glassman}.} \bibinfo{year}{2024}\natexlab{}.
\newblock \showarticletitle{Supporting Sensemaking of Large Language Model Outputs at Scale}. In \bibinfo{booktitle}{\emph{Proceedings of the CHI Conference on Human Factors in Computing Systems}} (Honolulu, HI, USA) \emph{(\bibinfo{series}{CHI '24})}. \bibinfo{publisher}{Association for Computing Machinery}, \bibinfo{address}{New York, NY, USA}, Article \bibinfo{articleno}{838}, \bibinfo{numpages}{21}~pages.
\newblock
\showISBNx{9798400703300}
\urldef\tempurl%
\url{https://doi.org/10.1145/3613904.3642139}
\showDOI{\tempurl}


\bibitem[Gonçalves et~al\mbox{.}(2013)]%
        {goncalves:ijdci:2013}
\bibfield{author}{\bibinfo{person}{Milene Gonçalves}, \bibinfo{person}{Carlos Cardoso}, {and} \bibinfo{person}{Petra Badke-Schaub}.} \bibinfo{year}{2013}\natexlab{}.
\newblock \showarticletitle{Inspiration peak: exploring the semantic distance between design problem and textual inspirational stimuli}.
\newblock \bibinfo{journal}{\emph{International Journal of Design Creativity and Innovation}} \bibinfo{volume}{1}, \bibinfo{number}{4} (\bibinfo{year}{2013}), \bibinfo{pages}{215--232}.
\newblock
\urldef\tempurl%
\url{https://doi.org/10.1080/21650349.2013.799309}
\showDOI{\tempurl}
\showeprint{https://doi.org/10.1080/21650349.2013.799309}


\bibitem[Grant and Berry(2011)]%
        {grant:2011:pf_stakeholder_perspective}
\bibfield{author}{\bibinfo{person}{Adam~M. Grant} {and} \bibinfo{person}{James~W. Berry}.} \bibinfo{year}{2011}\natexlab{}.
\newblock \showarticletitle{The Necessity of Others is The Mother of Invention: Intrinsic and Prosocial Motivations, Perspective Taking, and Creativity}.
\newblock \bibinfo{journal}{\emph{Academy of Management Journal}} \bibinfo{volume}{54}, \bibinfo{number}{1} (\bibinfo{year}{2011}), \bibinfo{pages}{73--96}.
\newblock
\urldef\tempurl%
\url{https://doi.org/10.5465/amj.2011.59215085}
\showDOI{\tempurl}
\showeprint{https://doi.org/10.5465/amj.2011.59215085}


\bibitem[Guo et~al\mbox{.}(2024)]%
        {guo:2024:agency_llm}
\bibfield{author}{\bibinfo{person}{Jiajing Guo}, \bibinfo{person}{Vikram Mohanty}, \bibinfo{person}{Jorge~H Piazentin~Ono}, \bibinfo{person}{Hongtao Hao}, \bibinfo{person}{Liang Gou}, {and} \bibinfo{person}{Liu Ren}.} \bibinfo{year}{2024}\natexlab{}.
\newblock \showarticletitle{Investigating Interaction Modes and User Agency in Human-LLM Collaboration for Domain-Specific Data Analysis}. In \bibinfo{booktitle}{\emph{Extended Abstracts of the 2024 CHI Conference on Human Factors in Computing Systems}} \emph{(\bibinfo{series}{CHI EA '24})}. \bibinfo{publisher}{Association for Computing Machinery}, \bibinfo{address}{New York, NY, USA}, Article \bibinfo{articleno}{203}, \bibinfo{numpages}{9}~pages.
\newblock
\showISBNx{9798400703317}
\urldef\tempurl%
\url{https://doi.org/10.1145/3613905.3651042}
\showDOI{\tempurl}


\bibitem[Haase and Laursen(2019)]%
        {haase:2019:pf_meaningFrames}
\bibfield{author}{\bibinfo{person}{Louise~Møller Haase} {and} \bibinfo{person}{Linda~Nhu Laursen}.} \bibinfo{year}{2019}\natexlab{}.
\newblock \showarticletitle{{Meaning Frames: The Structure of Problem Frames and Solution Frames}}.
\newblock \bibinfo{journal}{\emph{Design Issues}} \bibinfo{volume}{35}, \bibinfo{number}{3} (\bibinfo{date}{07} \bibinfo{year}{2019}), \bibinfo{pages}{20--34}.
\newblock
\showISSN{0747-9360}
\urldef\tempurl%
\url{https://doi.org/10.1162/desi_a_00547}
\showDOI{\tempurl}
\showeprint{https://direct.mit.edu/desi/article-pdf/35/3/20/1716070/desi\_a\_00547.pdf}


\bibitem[Halskov and Dalsg\r{a}rd(2006)]%
        {kim:dis:2006}
\bibfield{author}{\bibinfo{person}{Kim Halskov} {and} \bibinfo{person}{Peter Dalsg\r{a}rd}.} \bibinfo{year}{2006}\natexlab{}.
\newblock \showarticletitle{Inspiration Card Workshops}. In \bibinfo{booktitle}{\emph{Proceedings of the 6th Conference on Designing Interactive Systems}} (University Park, PA, USA) \emph{(\bibinfo{series}{DIS '06})}. \bibinfo{publisher}{Association for Computing Machinery}, \bibinfo{address}{New York, NY, USA}, \bibinfo{pages}{2–11}.
\newblock
\showISBNx{1595933670}
\urldef\tempurl%
\url{https://doi.org/10.1145/1142405.1142409}
\showDOI{\tempurl}


\bibitem[H\"{a}m\"{a}l\"{a}inen et~al\mbox{.}(2023)]%
        {perttu:chi:2023}
\bibfield{author}{\bibinfo{person}{Perttu H\"{a}m\"{a}l\"{a}inen}, \bibinfo{person}{Mikke Tavast}, {and} \bibinfo{person}{Anton Kunnari}.} \bibinfo{year}{2023}\natexlab{}.
\newblock \showarticletitle{Evaluating Large Language Models in Generating Synthetic HCI Research Data: A Case Study}. In \bibinfo{booktitle}{\emph{Proceedings of the 2023 CHI Conference on Human Factors in Computing Systems}} (Hamburg, Germany) \emph{(\bibinfo{series}{CHI '23})}. \bibinfo{publisher}{Association for Computing Machinery}, \bibinfo{address}{New York, NY, USA}, Article \bibinfo{articleno}{433}, \bibinfo{numpages}{19}~pages.
\newblock
\showISBNx{9781450394215}
\urldef\tempurl%
\url{https://doi.org/10.1145/3544548.3580688}
\showDOI{\tempurl}


\bibitem[Hart(2006)]%
        {hart:2006:nasa}
\bibfield{author}{\bibinfo{person}{Sandra~G Hart}.} \bibinfo{year}{2006}\natexlab{}.
\newblock \showarticletitle{NASA-task load index (NASA-TLX); 20 years later}. In \bibinfo{booktitle}{\emph{Proceedings of the human factors and ergonomics society annual meeting}}, Vol.~\bibinfo{volume}{50}. Sage publications Sage CA: Los Angeles, CA, \bibinfo{pages}{904--908}.
\newblock


\bibitem[He et~al\mbox{.}(2024)]%
        {he:2024:ai_groupIdeation}
\bibfield{author}{\bibinfo{person}{Jessica He}, \bibinfo{person}{Stephanie Houde}, \bibinfo{person}{Gabriel~E. Gonzalez}, \bibinfo{person}{Dar\'{\i}o~Andr\'{e}s Silva~Moran}, \bibinfo{person}{Steven~I. Ross}, \bibinfo{person}{Michael Muller}, {and} \bibinfo{person}{Justin~D. Weisz}.} \bibinfo{year}{2024}\natexlab{}.
\newblock \showarticletitle{AI and the Future of Collaborative Work: Group Ideation with an LLM in a Virtual Canvas}. In \bibinfo{booktitle}{\emph{Proceedings of the 3rd Annual Meeting of the Symposium on Human-Computer Interaction for Work}} (Newcastle upon Tyne, United Kingdom) \emph{(\bibinfo{series}{CHIWORK '24})}. \bibinfo{publisher}{Association for Computing Machinery}, \bibinfo{address}{New York, NY, USA}, Article \bibinfo{articleno}{9}, \bibinfo{numpages}{14}~pages.
\newblock
\showISBNx{9798400710179}
\urldef\tempurl%
\url{https://doi.org/10.1145/3663384.3663398}
\showDOI{\tempurl}


\bibitem[James Lloyd-Cox and Bhattacharya(2022)]%
        {james:2022:evaluating_creativity}
\bibfield{author}{\bibinfo{person}{Alan~Pickering James Lloyd-Cox} {and} \bibinfo{person}{Joydeep Bhattacharya}.} \bibinfo{year}{2022}\natexlab{}.
\newblock \showarticletitle{Evaluating Creativity: How Idea Context and Rater Personality Affect Considerations of Novelty and Usefulness}.
\newblock \bibinfo{journal}{\emph{Creativity Research Journal}} \bibinfo{volume}{34}, \bibinfo{number}{4} (\bibinfo{year}{2022}), \bibinfo{pages}{373--390}.
\newblock
\urldef\tempurl%
\url{https://doi.org/10.1080/10400419.2022.2125721}
\showDOI{\tempurl}
\showeprint{https://doi.org/10.1080/10400419.2022.2125721}


\bibitem[Jiang and Luo(2024)]%
        {jiang:2024:autotriz}
\bibfield{author}{\bibinfo{person}{Shuo Jiang} {and} \bibinfo{person}{Jianxi Luo}.} \bibinfo{year}{2024}\natexlab{}.
\newblock \bibinfo{title}{AutoTRIZ: Artificial Ideation with TRIZ and Large Language Models}.
\newblock
\newblock
\showeprint[arxiv]{2403.13002}
\urldef\tempurl%
\url{https://arxiv.org/abs/2403.13002}
\showURL{%
\tempurl}


\bibitem[Kim et~al\mbox{.}(2023)]%
        {llm_creativity:dis:2023}
\bibfield{author}{\bibinfo{person}{Jeongyeon Kim}, \bibinfo{person}{Sangho Suh}, \bibinfo{person}{Lydia~B Chilton}, {and} \bibinfo{person}{Haijun Xia}.} \bibinfo{year}{2023}\natexlab{}.
\newblock \showarticletitle{Metaphorian: Leveraging Large Language Models to Support Extended Metaphor Creation for Science Writing}. In \bibinfo{booktitle}{\emph{Proceedings of the 2023 ACM Designing Interactive Systems Conference}} \emph{(\bibinfo{series}{DIS '23})}. \bibinfo{publisher}{Association for Computing Machinery}, \bibinfo{address}{New York, NY, USA}, \bibinfo{pages}{115–135}.
\newblock
\showISBNx{9781450398930}
\urldef\tempurl%
\url{https://doi.org/10.1145/3563657.3595996}
\showDOI{\tempurl}


\bibitem[Klingbeil et~al\mbox{.}(2024)]%
        {klingbeil:2024:ai_overreliance}
\bibfield{author}{\bibinfo{person}{Artur Klingbeil}, \bibinfo{person}{Cassandra Grützner}, {and} \bibinfo{person}{Philipp Schreck}.} \bibinfo{year}{2024}\natexlab{}.
\newblock \showarticletitle{Trust and reliance on AI — An experimental study on the extent and costs of overreliance on AI}.
\newblock \bibinfo{journal}{\emph{Computers in Human Behavior}}  \bibinfo{volume}{160} (\bibinfo{year}{2024}), \bibinfo{pages}{108352}.
\newblock
\showISSN{0747-5632}
\urldef\tempurl%
\url{https://doi.org/10.1016/j.chb.2024.108352}
\showDOI{\tempurl}


\bibitem[Koivisto and Grassini(2023)]%
        {koivisto:2023:human_outperform_ai}
\bibfield{author}{\bibinfo{person}{Mika Koivisto} {and} \bibinfo{person}{Simone Grassini}.} \bibinfo{year}{2023}\natexlab{}.
\newblock \showarticletitle{Best humans still outperform artificial intelligence in a creative divergent thinking task}.
\newblock \bibinfo{journal}{\emph{Scientific reports}} \bibinfo{volume}{13}, \bibinfo{number}{1} (\bibinfo{year}{2023}), \bibinfo{pages}{13601}.
\newblock


\bibitem[Kr{\"o}per et~al\mbox{.}(2011)]%
        {kroper:2011:motivation_design_creativity}
\bibfield{author}{\bibinfo{person}{Madeleine Kr{\"o}per}, \bibinfo{person}{Doris Fay}, \bibinfo{person}{Tilmann Lindberg}, {and} \bibinfo{person}{Christoph Meinel}.} \bibinfo{year}{2011}\natexlab{}.
\newblock \showarticletitle{Interrelations between Motivation, Creativity and Emotions in Design Thinking Processes -- An Empirical Study Based on Regulatory Focus Theory}. In \bibinfo{booktitle}{\emph{Design Creativity 2010}}, \bibfield{editor}{\bibinfo{person}{Toshiharu Taura} {and} \bibinfo{person}{Yukari Nagai}} (Eds.). \bibinfo{publisher}{Springer London}, \bibinfo{address}{London}, \bibinfo{pages}{97--104}.
\newblock
\showISBNx{978-0-85729-224-7}


\bibitem[Lawton et~al\mbox{.}(2023)]%
        {lawton:2023:co-drawing_agency}
\bibfield{author}{\bibinfo{person}{Tomas Lawton}, \bibinfo{person}{Kazjon Grace}, {and} \bibinfo{person}{Francisco~J Ibarrola}.} \bibinfo{year}{2023}\natexlab{}.
\newblock \showarticletitle{When is a Tool a Tool? User Perceptions of System Agency in Human–AI Co-Creative Drawing}. In \bibinfo{booktitle}{\emph{Proceedings of the 2023 ACM Designing Interactive Systems Conference}} (Pittsburgh, PA, USA) \emph{(\bibinfo{series}{DIS '23})}. \bibinfo{publisher}{Association for Computing Machinery}, \bibinfo{address}{New York, NY, USA}, \bibinfo{pages}{1978–1996}.
\newblock
\showISBNx{9781450398930}
\urldef\tempurl%
\url{https://doi.org/10.1145/3563657.3595977}
\showDOI{\tempurl}


\bibitem[Lee(2020)]%
        {lee:2020:pf_convince}
\bibfield{author}{\bibinfo{person}{Jung-Joo Lee}.} \bibinfo{year}{2020}\natexlab{}.
\newblock \showarticletitle{Frame failures and reframing dialogues in the public sector design projects}.
\newblock \bibinfo{journal}{\emph{International Journal of Design}} \bibinfo{volume}{14}, \bibinfo{number}{1} (\bibinfo{year}{2020}), \bibinfo{pages}{81--94}.
\newblock


\bibitem[Li et~al\mbox{.}(2024a)]%
        {li:2024:ux_perception_ai}
\bibfield{author}{\bibinfo{person}{Jie Li}, \bibinfo{person}{Hancheng Cao}, \bibinfo{person}{Laura Lin}, \bibinfo{person}{Youyang Hou}, \bibinfo{person}{Ruihao Zhu}, {and} \bibinfo{person}{Abdallah El~Ali}.} \bibinfo{year}{2024}\natexlab{a}.
\newblock \showarticletitle{User Experience Design Professionals’ Perceptions of Generative Artificial Intelligence}. In \bibinfo{booktitle}{\emph{Proceedings of the CHI Conference on Human Factors in Computing Systems}} (Honolulu, HI, USA) \emph{(\bibinfo{series}{CHI '24})}. \bibinfo{publisher}{Association for Computing Machinery}, \bibinfo{address}{New York, NY, USA}, Article \bibinfo{articleno}{381}, \bibinfo{numpages}{18}~pages.
\newblock
\showISBNx{9798400703300}
\urldef\tempurl%
\url{https://doi.org/10.1145/3613904.3642114}
\showDOI{\tempurl}


\bibitem[Li et~al\mbox{.}(2024c)]%
        {li:2024:interaction_pattern_LLM}
\bibfield{author}{\bibinfo{person}{Jiayang Li}, \bibinfo{person}{Jiale Li}, {and} \bibinfo{person}{Yunsheng Su}.} \bibinfo{year}{2024}\natexlab{c}.
\newblock \showarticletitle{A Map of Exploring Human Interaction Patterns with LLM: Insights into Collaboration and Creativity}. In \bibinfo{booktitle}{\emph{Artificial Intelligence in HCI}}, \bibfield{editor}{\bibinfo{person}{Helmut Degen} {and} \bibinfo{person}{Stavroula Ntoa}} (Eds.). \bibinfo{publisher}{Springer Nature Switzerland}, \bibinfo{address}{Cham}, \bibinfo{pages}{60--85}.
\newblock
\showISBNx{978-3-031-60615-1}


\bibitem[Li et~al\mbox{.}(2024b)]%
        {li:2024:motivation_empathy_creativity}
\bibfield{author}{\bibinfo{person}{Xinyu Li}, \bibinfo{person}{Juanjuan Chen}, {and} \bibinfo{person}{Hongjie Fu}.} \bibinfo{year}{2024}\natexlab{b}.
\newblock \showarticletitle{The roles of empathy and motivation in creativity in design thinking}.
\newblock \bibinfo{journal}{\emph{International Journal of Technology and Design Education}} (\bibinfo{year}{2024}), \bibinfo{pages}{1--20}.
\newblock


\bibitem[Lin et~al\mbox{.}(2024)]%
        {lin:2024:ai_speech_writing}
\bibfield{author}{\bibinfo{person}{Susan Lin}, \bibinfo{person}{Jeremy Warner}, \bibinfo{person}{J.D. Zamfirescu-Pereira}, \bibinfo{person}{Matthew~G Lee}, \bibinfo{person}{Sauhard Jain}, \bibinfo{person}{Shanqing Cai}, \bibinfo{person}{Piyawat Lertvittayakumjorn}, \bibinfo{person}{Michael~Xuelin Huang}, \bibinfo{person}{Shumin Zhai}, \bibinfo{person}{Bjoern Hartmann}, {and} \bibinfo{person}{Can Liu}.} \bibinfo{year}{2024}\natexlab{}.
\newblock \showarticletitle{Rambler: Supporting Writing With Speech via LLM-Assisted Gist Manipulation}. In \bibinfo{booktitle}{\emph{Proceedings of the CHI Conference on Human Factors in Computing Systems}} (Honolulu, HI, USA) \emph{(\bibinfo{series}{CHI '24})}. \bibinfo{publisher}{Association for Computing Machinery}, \bibinfo{address}{New York, NY, USA}, Article \bibinfo{articleno}{1043}, \bibinfo{numpages}{19}~pages.
\newblock
\showISBNx{9798400703300}
\urldef\tempurl%
\url{https://doi.org/10.1145/3613904.3642217}
\showDOI{\tempurl}


\bibitem[Liu and Chilton(2022)]%
        {liu:2022:prompt_image_chart}
\bibfield{author}{\bibinfo{person}{Vivian Liu} {and} \bibinfo{person}{Lydia~B Chilton}.} \bibinfo{year}{2022}\natexlab{}.
\newblock \showarticletitle{Design Guidelines for Prompt Engineering Text-to-Image Generative Models}. In \bibinfo{booktitle}{\emph{Proceedings of the 2022 CHI Conference on Human Factors in Computing Systems}} (New Orleans, LA, USA) \emph{(\bibinfo{series}{CHI '22})}. \bibinfo{publisher}{Association for Computing Machinery}, \bibinfo{address}{New York, NY, USA}, Article \bibinfo{articleno}{384}, \bibinfo{numpages}{23}~pages.
\newblock
\showISBNx{9781450391573}
\urldef\tempurl%
\url{https://doi.org/10.1145/3491102.3501825}
\showDOI{\tempurl}


\bibitem[Long et~al\mbox{.}(2024)]%
        {long:2024:novelty_ai_workflow}
\bibfield{author}{\bibinfo{person}{Tao Long}, \bibinfo{person}{Katy~Ilonka Gero}, {and} \bibinfo{person}{Lydia~B Chilton}.} \bibinfo{year}{2024}\natexlab{}.
\newblock \showarticletitle{Not Just Novelty: A Longitudinal Study on Utility and Customization of an AI Workflow}. In \bibinfo{booktitle}{\emph{Proceedings of the 2024 ACM Designing Interactive Systems Conference}} (Copenhagen, Denmark) \emph{(\bibinfo{series}{DIS '24})}. \bibinfo{publisher}{Association for Computing Machinery}, \bibinfo{address}{New York, NY, USA}, \bibinfo{pages}{782–803}.
\newblock
\showISBNx{9798400705830}
\urldef\tempurl%
\url{https://doi.org/10.1145/3643834.3661587}
\showDOI{\tempurl}


\bibitem[Mensch and Gonçalves(2019)]%
        {Mensch:2019:pf_canvas}
\bibfield{author}{\bibinfo{person}{Sabine~Liana Mensch} {and} \bibinfo{person}{Milene Gonçalves}.} \bibinfo{year}{2019}\natexlab{}.
\newblock \showarticletitle{Tackling Reframing: The Development and Evaluation of a Problem Reframing Canvas}.
\newblock \bibinfo{journal}{\emph{Proceedings of the Design Society: International Conference on Engineering Design}} \bibinfo{volume}{1}, \bibinfo{number}{1} (\bibinfo{year}{2019}), \bibinfo{pages}{379–388}.
\newblock
\urldef\tempurl%
\url{https://doi.org/10.1017/dsi.2019.41}
\showDOI{\tempurl}


\bibitem[Mora et~al\mbox{.}(2017)]%
        {mora:2017:structured_ideation_tool}
\bibfield{author}{\bibinfo{person}{Simone Mora}, \bibinfo{person}{Francesco Gianni}, {and} \bibinfo{person}{Monica Divitini}.} \bibinfo{year}{2017}\natexlab{}.
\newblock \showarticletitle{Tiles: A Card-based Ideation Toolkit for the Internet of Things}. In \bibinfo{booktitle}{\emph{Proceedings of the 2017 Conference on Designing Interactive Systems}} (Edinburgh, United Kingdom) \emph{(\bibinfo{series}{DIS '17})}. \bibinfo{publisher}{Association for Computing Machinery}, \bibinfo{address}{New York, NY, USA}, \bibinfo{pages}{587–598}.
\newblock
\showISBNx{9781450349222}
\urldef\tempurl%
\url{https://doi.org/10.1145/3064663.3064699}
\showDOI{\tempurl}


\bibitem[Moruzzi(2022)]%
        {moruzzi:2022:AI_agency}
\bibfield{author}{\bibinfo{person}{Caterina Moruzzi}.} \bibinfo{year}{2022}\natexlab{}.
\newblock \showarticletitle{Creative Agents: Rethinking Agency and Creativity in Human and Artificial Systems}.
\newblock \bibinfo{journal}{\emph{Journal of Aesthetics and Phenomenology}} \bibinfo{volume}{9}, \bibinfo{number}{2} (\bibinfo{year}{2022}), \bibinfo{pages}{245--268}.
\newblock
\urldef\tempurl%
\url{https://doi.org/10.1080/20539320.2022.2150470}
\showDOI{\tempurl}
\showeprint{https://doi.org/10.1080/20539320.2022.2150470}


\bibitem[Noy and Zhang(2023)]%
        {noy:2023:llm_writing_good}
\bibfield{author}{\bibinfo{person}{Shakked Noy} {and} \bibinfo{person}{Whitney Zhang}.} \bibinfo{year}{2023}\natexlab{}.
\newblock \showarticletitle{Experimental evidence on the productivity effects of generative artificial intelligence}.
\newblock \bibinfo{journal}{\emph{Science}} \bibinfo{volume}{381}, \bibinfo{number}{6654} (\bibinfo{year}{2023}), \bibinfo{pages}{187--192}.
\newblock
\urldef\tempurl%
\url{https://doi.org/10.1126/science.adh2586}
\showDOI{\tempurl}
\showeprint{https://www.science.org/doi/pdf/10.1126/science.adh2586}


\bibitem[OpenAI(2023)]%
        {openai:2023:gpt4}
\bibfield{author}{\bibinfo{person}{OpenAI}.} \bibinfo{year}{2023}\natexlab{}.
\newblock \bibinfo{title}{GPT-4 Technical Report}.
\newblock
\newblock
\showeprint[arxiv]{2303.08774}


\bibitem[Passi and Vorvoreanu(2022)]%
        {passi:2022:ai_overreliance}
\bibfield{author}{\bibinfo{person}{Samir Passi} {and} \bibinfo{person}{Mihaela Vorvoreanu}.} \bibinfo{year}{2022}\natexlab{}.
\newblock \showarticletitle{Overreliance on AI literature review}.
\newblock \bibinfo{journal}{\emph{Microsoft Research}} (\bibinfo{year}{2022}).
\newblock


\bibitem[Paton and Dorst(2011)]%
        {paton:2011:briefing_reframing}
\bibfield{author}{\bibinfo{person}{Bec Paton} {and} \bibinfo{person}{Kees Dorst}.} \bibinfo{year}{2011}\natexlab{}.
\newblock \showarticletitle{Briefing and reframing: A situated practice}.
\newblock \bibinfo{journal}{\emph{Design Studies}} \bibinfo{volume}{32}, \bibinfo{number}{6} (\bibinfo{year}{2011}), \bibinfo{pages}{573--587}.
\newblock
\showISSN{0142-694X}
\urldef\tempurl%
\url{https://doi.org/10.1016/j.destud.2011.07.002}
\showDOI{\tempurl}
\newblock
\shownote{Interpreting Design Thinking}.


\bibitem[Paulus et~al\mbox{.}(2011)]%
        {paulus:jcb:2011}
\bibfield{author}{\bibinfo{person}{Paul~B Paulus}, \bibinfo{person}{Nicholas~W Kohn}, {and} \bibinfo{person}{Lauren~E Arditti}.} \bibinfo{year}{2011}\natexlab{}.
\newblock \showarticletitle{Effects of Quantity and Quality Instructions on Brainstorming}.
\newblock \bibinfo{journal}{\emph{The Journal of Creative Behavior}} \bibinfo{volume}{45}, \bibinfo{number}{1} (\bibinfo{year}{2011}), \bibinfo{pages}{38--46}.
\newblock
\urldef\tempurl%
\url{https://doi.org/10.1002/j.2162-6057.2011.tb01083.x}
\showDOI{\tempurl}
\showeprint{https://onlinelibrary.wiley.com/doi/pdf/10.1002/j.2162-6057.2011.tb01083.x}


\bibitem[Pee et~al\mbox{.}(2015)]%
        {pee:2015:pf_seeing_different}
\bibfield{author}{\bibinfo{person}{Suat~Hoon Pee}, \bibinfo{person}{CH Dorst}, {and} \bibinfo{person}{Mieke van~der Bijl-Brouwer}.} \bibinfo{year}{2015}\natexlab{}.
\newblock \showarticletitle{Understanding problem framing through research into metaphors}. In \bibinfo{booktitle}{\emph{2015 IASDR International Design Research Conference}}. \bibinfo{pages}{1656--1671}.
\newblock


\bibitem[Qin et~al\mbox{.}(2024)]%
        {qin:2024:ai_character_writing}
\bibfield{author}{\bibinfo{person}{Hua~Xuan Qin}, \bibinfo{person}{Shan Jin}, \bibinfo{person}{Ze Gao}, \bibinfo{person}{Mingming Fan}, {and} \bibinfo{person}{Pan Hui}.} \bibinfo{year}{2024}\natexlab{}.
\newblock \showarticletitle{CharacterMeet: Supporting Creative Writers' Entire Story Character Construction Processes Through Conversation with LLM-Powered Chatbot Avatars}. In \bibinfo{booktitle}{\emph{Proceedings of the CHI Conference on Human Factors in Computing Systems}} (Honolulu, HI, USA) \emph{(\bibinfo{series}{CHI '24})}. \bibinfo{publisher}{Association for Computing Machinery}, \bibinfo{address}{New York, NY, USA}, Article \bibinfo{articleno}{1051}, \bibinfo{numpages}{19}~pages.
\newblock
\showISBNx{9798400703300}
\urldef\tempurl%
\url{https://doi.org/10.1145/3613904.3642105}
\showDOI{\tempurl}


\bibitem[Runco and Jaeger(2012)]%
        {runco:2012:creativity}
\bibfield{author}{\bibinfo{person}{Mark~A. Runco} {and} \bibinfo{person}{Garrett~J. Jaeger}.} \bibinfo{year}{2012}\natexlab{}.
\newblock \showarticletitle{The Standard Definition of Creativity}.
\newblock \bibinfo{journal}{\emph{Creativity Research Journal}} \bibinfo{volume}{24}, \bibinfo{number}{1} (\bibinfo{year}{2012}), \bibinfo{pages}{92--96}.
\newblock
\urldef\tempurl%
\url{https://doi.org/10.1080/10400419.2012.650092}
\showDOI{\tempurl}
\showeprint{https://doi.org/10.1080/10400419.2012.650092}


\bibitem[Sandholm et~al\mbox{.}(2024)]%
        {sandholm:2024:llm_randomness}
\bibfield{author}{\bibinfo{person}{Thomas Sandholm}, \bibinfo{person}{Sayandev Mukherjee}, {and} \bibinfo{person}{Bernardo~A. Huberman}.} \bibinfo{year}{2024}\natexlab{}.
\newblock \bibinfo{title}{Randomness Is All You Need: Semantic Traversal of Problem-Solution Spaces with Large Language Models}.
\newblock
\newblock
\showeprint[arxiv]{2402.06053}
\urldef\tempurl%
\url{https://arxiv.org/abs/2402.06053}
\showURL{%
\tempurl}


\bibitem[Sarkar and Chakrabarti(2011)]%
        {sarkar:2011:assessing_design_creativity}
\bibfield{author}{\bibinfo{person}{Prabir Sarkar} {and} \bibinfo{person}{Amaresh Chakrabarti}.} \bibinfo{year}{2011}\natexlab{}.
\newblock \showarticletitle{Assessing design creativity}.
\newblock \bibinfo{journal}{\emph{Design Studies}} \bibinfo{volume}{32}, \bibinfo{number}{4} (\bibinfo{year}{2011}), \bibinfo{pages}{348--383}.
\newblock
\showISSN{0142-694X}
\urldef\tempurl%
\url{https://doi.org/10.1016/j.destud.2011.01.002}
\showDOI{\tempurl}


\bibitem[Schmidt et~al\mbox{.}(2024)]%
        {schmidt:2024:hcd-chatgpt}
\bibfield{author}{\bibinfo{person}{Albrecht Schmidt}, \bibinfo{person}{Passant Elagroudy}, \bibinfo{person}{Fiona Draxler}, \bibinfo{person}{Frauke Kreuter}, {and} \bibinfo{person}{Robin Welsch}.} \bibinfo{year}{2024}\natexlab{}.
\newblock \showarticletitle{Simulating the Human in HCD with ChatGPT: Redesigning Interaction Design with AI}.
\newblock \bibinfo{journal}{\emph{Interactions}} \bibinfo{volume}{31}, \bibinfo{number}{1} (\bibinfo{date}{jan} \bibinfo{year}{2024}), \bibinfo{pages}{24–31}.
\newblock
\showISSN{1072-5520}
\urldef\tempurl%
\url{https://doi.org/10.1145/3637436}
\showDOI{\tempurl}


\bibitem[Sch{\"o}n(2017)]%
        {schon:book:reflective}
\bibfield{author}{\bibinfo{person}{Donald~A Sch{\"o}n}.} \bibinfo{year}{2017}\natexlab{}.
\newblock \bibinfo{booktitle}{\emph{The reflective practitioner: How professionals think in action}}.
\newblock \bibinfo{publisher}{Routledge}.
\newblock


\bibitem[Segal(2004)]%
        {segal:2004:creativity_incubation}
\bibfield{author}{\bibinfo{person}{Eliaz Segal}.} \bibinfo{year}{2004}\natexlab{}.
\newblock \showarticletitle{Incubation in Insight Problem Solving}.
\newblock \bibinfo{journal}{\emph{Creativity Research Journal}} \bibinfo{volume}{16}, \bibinfo{number}{1} (\bibinfo{year}{2004}), \bibinfo{pages}{141--148}.
\newblock
\urldef\tempurl%
\url{https://doi.org/10.1207/s15326934crj1601\_13}
\showDOI{\tempurl}
\showeprint{https://doi.org/10.1207/s15326934crj1601\_13}


\bibitem[Shaer et~al\mbox{.}(2024)]%
        {share:2024:brainwriting_llm}
\bibfield{author}{\bibinfo{person}{Orit Shaer}, \bibinfo{person}{Angelora Cooper}, \bibinfo{person}{Osnat Mokryn}, \bibinfo{person}{Andrew~L Kun}, {and} \bibinfo{person}{Hagit Ben~Shoshan}.} \bibinfo{year}{2024}\natexlab{}.
\newblock \showarticletitle{AI-Augmented Brainwriting: Investigating the use of LLMs in group ideation}. In \bibinfo{booktitle}{\emph{Proceedings of the CHI Conference on Human Factors in Computing Systems}} (Honolulu, HI, USA) \emph{(\bibinfo{series}{CHI '24})}. \bibinfo{publisher}{Association for Computing Machinery}, \bibinfo{address}{New York, NY, USA}, Article \bibinfo{articleno}{1050}, \bibinfo{numpages}{17}~pages.
\newblock
\showISBNx{9798400703300}
\urldef\tempurl%
\url{https://doi.org/10.1145/3613904.3642414}
\showDOI{\tempurl}


\bibitem[Shanahan et~al\mbox{.}(2023)]%
        {roleplay/Shanahan2023}
\bibfield{author}{\bibinfo{person}{Murray Shanahan}, \bibinfo{person}{Kyle McDonell}, {and} \bibinfo{person}{Laria Reynolds}.} \bibinfo{year}{2023}\natexlab{}.
\newblock \showarticletitle{Role play with large language models}.
\newblock \bibinfo{journal}{\emph{Nature}} \bibinfo{volume}{623}, \bibinfo{number}{7987} (\bibinfo{date}{01 Nov} \bibinfo{year}{2023}), \bibinfo{pages}{493--498}.
\newblock
\showISSN{1476-4687}
\urldef\tempurl%
\url{https://doi.org/10.1038/s41586-023-06647-8}
\showDOI{\tempurl}


\bibitem[Shin et~al\mbox{.}(2024)]%
        {shin:2024:persona}
\bibfield{author}{\bibinfo{person}{Joongi Shin}, \bibinfo{person}{Michael~A. Hedderich}, \bibinfo{person}{Bart\l{}omiej~Jakub Rey}, \bibinfo{person}{Andr\'{e}s Lucero}, {and} \bibinfo{person}{Antti Oulasvirta}.} \bibinfo{year}{2024}\natexlab{}.
\newblock \showarticletitle{Understanding Human-AI Workflows for Generating Personas}. In \bibinfo{booktitle}{\emph{Proceedings of the 2024 ACM Designing Interactive Systems Conference}} (Copenhagen, Denmark) \emph{(\bibinfo{series}{DIS '24})}. \bibinfo{publisher}{Association for Computing Machinery}, \bibinfo{address}{New York, NY, USA}, \bibinfo{pages}{757–781}.
\newblock
\showISBNx{9798400705830}
\urldef\tempurl%
\url{https://doi.org/10.1145/3643834.3660729}
\showDOI{\tempurl}


\bibitem[Shin et~al\mbox{.}(2023)]%
        {shin:2023:human-ai}
\bibfield{author}{\bibinfo{person}{Joon~Gi Shin}, \bibinfo{person}{Janin Koch}, \bibinfo{person}{Andr\'{e}s Lucero}, \bibinfo{person}{Peter Dalsgaard}, {and} \bibinfo{person}{Wendy~E. Mackay}.} \bibinfo{year}{2023}\natexlab{}.
\newblock \showarticletitle{Integrating AI in Human-Human Collaborative Ideation}. In \bibinfo{booktitle}{\emph{Extended Abstracts of the 2023 CHI Conference on Human Factors in Computing Systems}} (Hamburg, Germany) \emph{(\bibinfo{series}{CHI EA '23})}. \bibinfo{publisher}{Association for Computing Machinery}, \bibinfo{address}{New York, NY, USA}, Article \bibinfo{articleno}{355}, \bibinfo{numpages}{5}~pages.
\newblock
\showISBNx{9781450394222}
\urldef\tempurl%
\url{https://doi.org/10.1145/3544549.3573802}
\showDOI{\tempurl}


\bibitem[Si et~al\mbox{.}(2024)]%
        {si:2024:llm_novel_research}
\bibfield{author}{\bibinfo{person}{Chenglei Si}, \bibinfo{person}{Diyi Yang}, {and} \bibinfo{person}{Tatsunori Hashimoto}.} \bibinfo{year}{2024}\natexlab{}.
\newblock \bibinfo{title}{Can LLMs Generate Novel Research Ideas? A Large-Scale Human Study with 100+ NLP Researchers}.
\newblock
\newblock
\showeprint[arxiv]{2409.04109}
\urldef\tempurl%
\url{https://arxiv.org/abs/2409.04109}
\showURL{%
\tempurl}


\bibitem[Siangliulue et~al\mbox{.}(2015)]%
        {siangliulue:cscw:2015}
\bibfield{author}{\bibinfo{person}{Pao Siangliulue}, \bibinfo{person}{Kenneth~C. Arnold}, \bibinfo{person}{Krzysztof~Z. Gajos}, {and} \bibinfo{person}{Steven~P. Dow}.} \bibinfo{year}{2015}\natexlab{}.
\newblock \showarticletitle{Toward Collaborative Ideation at Scale: Leveraging Ideas from Others to Generate More Creative and Diverse Ideas}. In \bibinfo{booktitle}{\emph{Proceedings of the 18th ACM Conference on Computer Supported Cooperative Work \&amp; Social Computing}} (Vancouver, BC, Canada) \emph{(\bibinfo{series}{CSCW '15})}. \bibinfo{publisher}{Association for Computing Machinery}, \bibinfo{address}{New York, NY, USA}, \bibinfo{pages}{937–945}.
\newblock
\showISBNx{9781450329224}
\urldef\tempurl%
\url{https://doi.org/10.1145/2675133.2675239}
\showDOI{\tempurl}


\bibitem[Silk et~al\mbox{.}(2021)]%
        {silk:pf_cognitiveStyle}
\bibfield{author}{\bibinfo{person}{Eli~M. Silk}, \bibinfo{person}{Amy~E. Rechkemmer}, \bibinfo{person}{Shanna~R. Daly}, \bibinfo{person}{Kathryn~W. Jablokow}, {and} \bibinfo{person}{Seda McKilligan}.} \bibinfo{year}{2021}\natexlab{}.
\newblock \showarticletitle{Problem framing and cognitive style: Impacts on design ideation perceptions}.
\newblock \bibinfo{journal}{\emph{Design Studies}}  \bibinfo{volume}{74} (\bibinfo{year}{2021}), \bibinfo{pages}{101015}.
\newblock
\showISSN{0142-694X}
\urldef\tempurl%
\url{https://doi.org/10.1016/j.destud.2021.101015}
\showDOI{\tempurl}


\bibitem[Simmons(2023)]%
        {prompt/simmons2023moral}
\bibfield{author}{\bibinfo{person}{Gabriel Simmons}.} \bibinfo{year}{2023}\natexlab{}.
\newblock \bibinfo{title}{Moral Mimicry: Large Language Models Produce Moral Rationalizations Tailored to Political Identity}.
\newblock
\newblock
\showeprint[arxiv]{2209.12106}


\bibitem[Simon(1995)]%
        {simon:1995:problemSolving}
\bibfield{author}{\bibinfo{person}{Herbert~A Simon}.} \bibinfo{year}{1995}\natexlab{}.
\newblock \showarticletitle{Problem forming, problem finding and problem solving in design}.
\newblock \bibinfo{journal}{\emph{Design \& systems}} (\bibinfo{year}{1995}), \bibinfo{pages}{245--257}.
\newblock


\bibitem[Smith and Blankenship(1991)]%
        {smith:1991:problem-solving_incubation}
\bibfield{author}{\bibinfo{person}{Steven~M. Smith} {and} \bibinfo{person}{Steven~E. Blankenship}.} \bibinfo{year}{1991}\natexlab{}.
\newblock \showarticletitle{Incubation and the Persistence of Fixation in Problem Solving}.
\newblock \bibinfo{journal}{\emph{The American Journal of Psychology}} \bibinfo{volume}{104}, \bibinfo{number}{1} (\bibinfo{year}{1991}), \bibinfo{pages}{61--87}.
\newblock
\showISSN{00029556}
\urldef\tempurl%
\url{http://www.jstor.org/stable/1422851}
\showURL{%
\tempurl}


\bibitem[Stompff et~al\mbox{.}(2016)]%
        {stompff:2016:pf_surprise}
\bibfield{author}{\bibinfo{person}{Guido Stompff}, \bibinfo{person}{Frido Smulders}, {and} \bibinfo{person}{Lilian Henze}.} \bibinfo{year}{2016}\natexlab{}.
\newblock \showarticletitle{Surprises are the benefits: reframing in multidisciplinary design teams}.
\newblock \bibinfo{journal}{\emph{Design Studies}}  \bibinfo{volume}{47} (\bibinfo{year}{2016}), \bibinfo{pages}{187--214}.
\newblock
\showISSN{0142-694X}
\urldef\tempurl%
\url{https://doi.org/10.1016/j.destud.2016.09.004}
\showDOI{\tempurl}


\bibitem[Sturdee et~al\mbox{.}(2015)]%
        {sturdee:2015:structured_ideation}
\bibfield{author}{\bibinfo{person}{Miriam Sturdee}, \bibinfo{person}{John Hardy}, \bibinfo{person}{Nick Dunn}, {and} \bibinfo{person}{Jason Alexander}.} \bibinfo{year}{2015}\natexlab{}.
\newblock \showarticletitle{A Public Ideation of Shape-Changing Applications}. In \bibinfo{booktitle}{\emph{Proceedings of the 2015 International Conference on Interactive Tabletops \& Surfaces}} (Madeira, Portugal) \emph{(\bibinfo{series}{ITS '15})}. \bibinfo{publisher}{Association for Computing Machinery}, \bibinfo{address}{New York, NY, USA}, \bibinfo{pages}{219–228}.
\newblock
\showISBNx{9781450338998}
\urldef\tempurl%
\url{https://doi.org/10.1145/2817721.2817734}
\showDOI{\tempurl}


\bibitem[Suh et~al\mbox{.}(2024)]%
        {suh:2024:luminate}
\bibfield{author}{\bibinfo{person}{Sangho Suh}, \bibinfo{person}{Meng Chen}, \bibinfo{person}{Bryan Min}, \bibinfo{person}{Toby Jia-Jun Li}, {and} \bibinfo{person}{Haijun Xia}.} \bibinfo{year}{2024}\natexlab{}.
\newblock \showarticletitle{Luminate: Structured Generation and Exploration of Design Space with Large Language Models for Human-AI Co-Creation}. In \bibinfo{booktitle}{\emph{Proceedings of the CHI Conference on Human Factors in Computing Systems}} (Honolulu, HI, USA) \emph{(\bibinfo{series}{CHI '24})}. \bibinfo{publisher}{Association for Computing Machinery}, \bibinfo{address}{New York, NY, USA}, Article \bibinfo{articleno}{644}, \bibinfo{numpages}{26}~pages.
\newblock
\showISBNx{9798400703300}
\urldef\tempurl%
\url{https://doi.org/10.1145/3613904.3642400}
\showDOI{\tempurl}


\bibitem[Suh et~al\mbox{.}(2023)]%
        {suh:2023:llm_sensemaking}
\bibfield{author}{\bibinfo{person}{Sangho Suh}, \bibinfo{person}{Bryan Min}, \bibinfo{person}{Srishti Palani}, {and} \bibinfo{person}{Haijun Xia}.} \bibinfo{year}{2023}\natexlab{}.
\newblock \showarticletitle{Sensecape: Enabling Multilevel Exploration and Sensemaking with Large Language Models}. In \bibinfo{booktitle}{\emph{Proceedings of the 36th Annual ACM Symposium on User Interface Software and Technology}} (San Francisco, CA, USA) \emph{(\bibinfo{series}{UIST '23})}. \bibinfo{publisher}{Association for Computing Machinery}, \bibinfo{address}{New York, NY, USA}, Article \bibinfo{articleno}{1}, \bibinfo{numpages}{18}~pages.
\newblock
\showISBNx{9798400701320}
\urldef\tempurl%
\url{https://doi.org/10.1145/3586183.3606756}
\showDOI{\tempurl}


\bibitem[Wadinambiarachchi et~al\mbox{.}(2024)]%
        {wadi:2024:Genai_creativity_fixation}
\bibfield{author}{\bibinfo{person}{Samangi Wadinambiarachchi}, \bibinfo{person}{Ryan~M. Kelly}, \bibinfo{person}{Saumya Pareek}, \bibinfo{person}{Qiushi Zhou}, {and} \bibinfo{person}{Eduardo Velloso}.} \bibinfo{year}{2024}\natexlab{}.
\newblock \showarticletitle{The Effects of Generative AI on Design Fixation and Divergent Thinking}. In \bibinfo{booktitle}{\emph{Proceedings of the CHI Conference on Human Factors in Computing Systems}} (Honolulu, HI, USA) \emph{(\bibinfo{series}{CHI '24})}. \bibinfo{publisher}{Association for Computing Machinery}, \bibinfo{address}{New York, NY, USA}, Article \bibinfo{articleno}{380}, \bibinfo{numpages}{18}~pages.
\newblock
\showISBNx{9798400703300}
\urldef\tempurl%
\url{https://doi.org/10.1145/3613904.3642919}
\showDOI{\tempurl}


\bibitem[Wen and Imamizu(2022)]%
        {wen:2022:sense_of_agency}
\bibfield{author}{\bibinfo{person}{Wen Wen} {and} \bibinfo{person}{Hiroshi Imamizu}.} \bibinfo{year}{2022}\natexlab{}.
\newblock \showarticletitle{The sense of agency in perception, behaviour and human--machine interactions}.
\newblock \bibinfo{journal}{\emph{Nature Reviews Psychology}} \bibinfo{volume}{1}, \bibinfo{number}{4} (\bibinfo{year}{2022}), \bibinfo{pages}{211--222}.
\newblock


\bibitem[White et~al\mbox{.}(2023)]%
        {white:arxiv:2023:prompt}
\bibfield{author}{\bibinfo{person}{Jules White}, \bibinfo{person}{Quchen Fu}, \bibinfo{person}{Sam Hays}, \bibinfo{person}{Michael Sandborn}, \bibinfo{person}{Carlos Olea}, \bibinfo{person}{Henry Gilbert}, \bibinfo{person}{Ashraf Elnashar}, \bibinfo{person}{Jesse Spencer-Smith}, {and} \bibinfo{person}{Douglas~C. Schmidt}.} \bibinfo{year}{2023}\natexlab{}.
\newblock \bibinfo{title}{A Prompt Pattern Catalog to Enhance Prompt Engineering with ChatGPT}.
\newblock
\newblock
\showeprint[arxiv]{2302.11382}


\bibitem[Wu et~al\mbox{.}(2022)]%
        {promptChaining:wu:2022}
\bibfield{author}{\bibinfo{person}{Tongshuang Wu}, \bibinfo{person}{Michael Terry}, {and} \bibinfo{person}{Carrie~Jun Cai}.} \bibinfo{year}{2022}\natexlab{}.
\newblock \showarticletitle{AI Chains: Transparent and Controllable Human-AI Interaction by Chaining Large Language Model Prompts}. In \bibinfo{booktitle}{\emph{Proceedings of the 2022 CHI Conference on Human Factors in Computing Systems}} \emph{(\bibinfo{series}{CHI '22})}. \bibinfo{publisher}{Association for Computing Machinery}, \bibinfo{address}{New York, NY, USA}, Article \bibinfo{articleno}{385}, \bibinfo{numpages}{22}~pages.
\newblock
\showISBNx{9781450391573}
\urldef\tempurl%
\url{https://doi.org/10.1145/3491102.3517582}
\showDOI{\tempurl}


\bibitem[Xu et~al\mbox{.}(2024)]%
        {xu:2024:llm_japmlate_why}
\bibfield{author}{\bibinfo{person}{Xiaotong~(Tone) Xu}, \bibinfo{person}{Jiayu Yin}, \bibinfo{person}{Catherine Gu}, \bibinfo{person}{Jenny Mar}, \bibinfo{person}{Sydney Zhang}, \bibinfo{person}{Jane~L. E}, {and} \bibinfo{person}{Steven~P. Dow}.} \bibinfo{year}{2024}\natexlab{}.
\newblock \showarticletitle{Jamplate: Exploring LLM-Enhanced Templates for Idea Reflection}. In \bibinfo{booktitle}{\emph{Proceedings of the 29th International Conference on Intelligent User Interfaces}} (Greenville, SC, USA) \emph{(\bibinfo{series}{IUI '24})}. \bibinfo{publisher}{Association for Computing Machinery}, \bibinfo{address}{New York, NY, USA}, \bibinfo{pages}{907–921}.
\newblock
\showISBNx{9798400705083}
\urldef\tempurl%
\url{https://doi.org/10.1145/3640543.3645196}
\showDOI{\tempurl}


\bibitem[Yuan et~al\mbox{.}(2022)]%
        {yuan:2022:wordcraft_writing}
\bibfield{author}{\bibinfo{person}{Ann Yuan}, \bibinfo{person}{Andy Coenen}, \bibinfo{person}{Emily Reif}, {and} \bibinfo{person}{Daphne Ippolito}.} \bibinfo{year}{2022}\natexlab{}.
\newblock \showarticletitle{Wordcraft: Story Writing With Large Language Models}. In \bibinfo{booktitle}{\emph{Proceedings of the 27th International Conference on Intelligent User Interfaces}} (Helsinki, Finland) \emph{(\bibinfo{series}{IUI '22})}. \bibinfo{publisher}{Association for Computing Machinery}, \bibinfo{address}{New York, NY, USA}, \bibinfo{pages}{841–852}.
\newblock
\showISBNx{9781450391443}
\urldef\tempurl%
\url{https://doi.org/10.1145/3490099.3511105}
\showDOI{\tempurl}


\bibitem[Zamfirescu-Pereira et~al\mbox{.}(2023)]%
        {zamfirescu:chi:2023}
\bibfield{author}{\bibinfo{person}{J.D. Zamfirescu-Pereira}, \bibinfo{person}{Richmond~Y. Wong}, \bibinfo{person}{Bjoern Hartmann}, {and} \bibinfo{person}{Qian Yang}.} \bibinfo{year}{2023}\natexlab{}.
\newblock \showarticletitle{Why Johnny Can’t Prompt: How Non-AI Experts Try (and Fail) to Design LLM Prompts}. In \bibinfo{booktitle}{\emph{Proceedings of the 2023 CHI Conference on Human Factors in Computing Systems}} (Hamburg, Germany) \emph{(\bibinfo{series}{CHI '23})}. \bibinfo{publisher}{Association for Computing Machinery}, \bibinfo{address}{New York, NY, USA}, Article \bibinfo{articleno}{437}, \bibinfo{numpages}{21}~pages.
\newblock
\showISBNx{9781450394215}
\urldef\tempurl%
\url{https://doi.org/10.1145/3544548.3581388}
\showDOI{\tempurl}


\end{thebibliography}


\raggedbottom

\pagebreak

\appendix
\onecolumn
\section*{{APPENDICES}}
\section{Implementation of the LLM-based Approaches}
\label{appendix:system}

We designed prompts and user interfaces to implement the \textsc{direct} and \textsc{structured} approaches.
For our study, it was crucial that we test our hypotheses with the highest-quality problem frames and intermediate content that LLMs can generate.
To this end, we iteratively designed prompts informed by guidelines for effective prompting \cite{white:arxiv:2023:prompt, Wynter:arxiv:2023, prompt/cox2023prompting, prompt/simmons2023moral}.
The main practice involved was to give the LLMs clear, concise task instructions while making sure to break complex tasks into smaller sequential subtasks that can reap the most from LLMs' reasoning competence.
Accordingly, we set up a system role that describes problem reframing and upcoming tasks (see Table \ref{tab:system_prompt}), then tested our prompts by using the model that performed best at the time of the research (GPT-4o). 

Another important factor we considered is how to guide users to interact with the LLMs as intended under each approach. 
While users converse freely with the agent in the \textsc{free-form} setting, the \textsc{direct} and \textsc{structured} ones should guide users to employ particular methods with the models.
Therefore, we designed a user interface wherein users click buttons to prompt the LLMs in each step of problem reframing.
This button-based interaction has the advantage of preventing users from putting the LLMs to different purposes; guaranteeing a similar experience across participants who differed in their prompting skills let us accurately assess the potential of each approach relative to others.
Below, we introduce the prompts and system design of the \textsc{direct} and the \textsc{structured} approach.

\begin{table}[h]
\centering
\small
\caption{The system prompt shared by the \textsc{direct} and the \textsc{structured} approach}
\begin{tabularx}{\textwidth}{lX}
\toprule
\textbf{Role} & \textbf{Content} \\
\midrule
System & You are an expert designer performing "problem reframing". Problem reframing is an essential process of solving design problems, where designers explore alternative ways of defining initial problems that can lead to creative and actionable solutions. For this, you will be given descriptions related to problems and a task for reframing the problem. \\
\bottomrule
\end{tabularx}
\label{tab:system_prompt}
\end{table}

\subsection{The Direct Approach}
\label{appendix:system_direct}
\begin{figure*}[h]
    \centering
    \small
    \includegraphics[width=\linewidth]{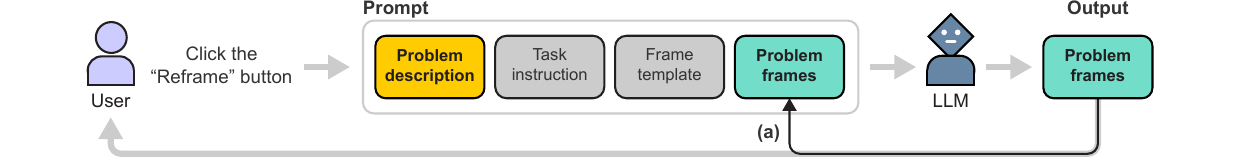}
    \caption{The \textsc{direct} approach employs LLM-generated problem frames to yield a broader array of problem frames in the next run (\texttt{a}).}
    \Description{}
    \label{fig:direct_diagram}
\end{figure*}

\begin{figure*}[t]
    \centering
    \includegraphics[width=\linewidth]{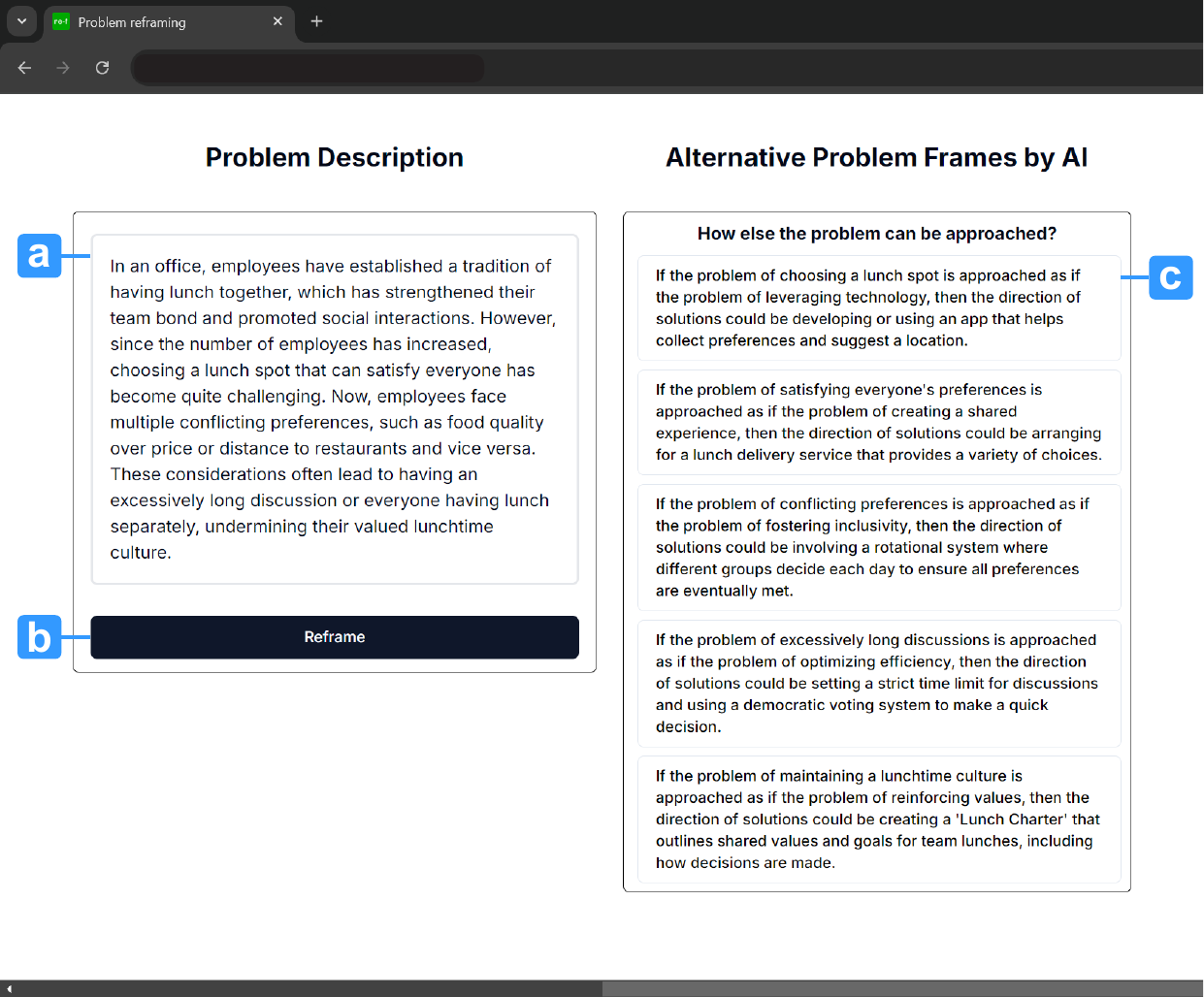}
    \caption{The interface for the \textsc{direct} approach. On the left is the description of the problem that users need to reframe (\texttt{a}). Pressing the ``Reframe'' button (\texttt{b}) causes the LLMs to generate alternative problem frames, presented in one frame per row (\texttt{c}).}
    \Description{}
    \label{fig:direct}
\end{figure*}

The core idea behind the \textsc{direct} approach is that designers interact with LLM-generated problem frames only (Figure \ref{fig:direct_diagram} presents this principle).
To enable that, we designed an interface in which users click a button to prompt LLMs to generate alternative problem frames. In this interface, presented in Figure \ref{fig:direct}, 
users read a problem description, press the button to (re)generate alternative problem frames, and build on the LLM-generated frames.
To make the process iterative, we designed two prompts each comprising input, task, and output components (shown in Table \ref{tab:direct_prompt}).
The first prompt is used to generate the initial set of frames on the basis of the problem description alone.
This prompt includes a problem description, a reframing task, and a format for describing problem frames (following Dorst's frame template \cite{dorst:book:reframing}).
The second prompt is used iteratively. Here, the previous frames function as input to instructing LLMs to generate different frames.

\begin{table}[t]
\centering
\small
\caption{Prompts for the \textsc{direct} approach}
\begin{tabularx}{\textwidth}{p{0.25\textwidth}X}
\toprule
\textbf{Generating initial frames} &
Here is a description of a problem situation: \texttt{\{problem description\}}\newline
Generate alternative ways of approaching the problem.\newline
Show only the result in the following format: \texttt{\{problem frame template\}}\\
\midrule
\textbf{Generating alternative frames} &
Here is a description of a problem situation: \texttt{\{problem description\}}\newline
Here are alternative ways of approaching the problem: \texttt{\{previous problem frames\}}\newline
Compared to the approaches above, generate different ways of approaching the problem.\newline
Show only the result in the following format: \texttt{\{problem frame template\}}
\\
\bottomrule
\end{tabularx}
\label{tab:direct_prompt}
\end{table}

\clearpage

\subsection{The Structured Approach}
\label{appendix:system_structure}
\begin{figure*}[h]
    \centering
    \small
    \includegraphics[width=\linewidth]{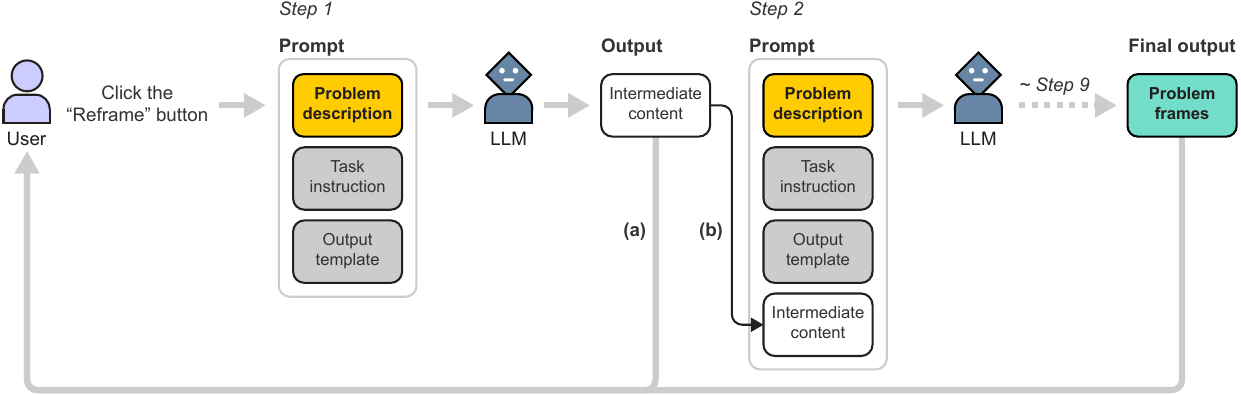}
    \caption{In the \textsc{structured} approach, LLMs generate alternative frames by following Dorst's nine-step reframing process. Throughout the process, users can review and generate new specific content (\texttt{a}), which is used to create the content in the following steps (\texttt{b}).}
    \Description{}
    \label{fig:structured_diagram}
\end{figure*}

\begin{figure*}[t]
    \centering
    \small
    \includegraphics[width=\linewidth]{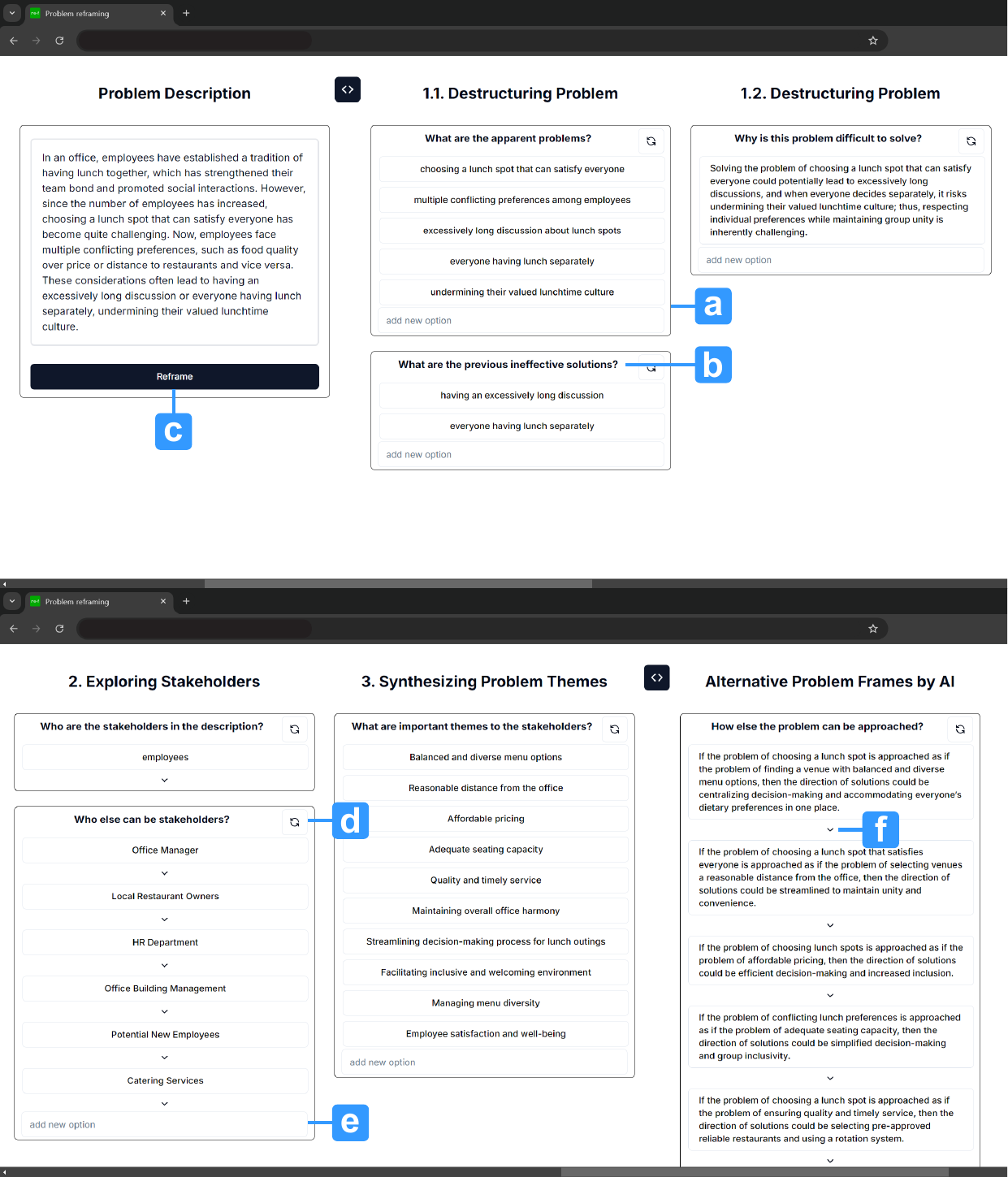}
    \caption{The interface for the \textsc{structured} approach. The leftmost of its six columns displays a problem description while the others show LLM-generated content in every successive window (\texttt{a}) to represent each step in Dorst's reframing process, in turn. So that the nature of the content is clear at all times, the windows display the key points in each step (\texttt{b}). Users can click ``Reframe'' to prompt the LLMs to run through the entire set of steps (\texttt{c}). They can also click the re-generation button, to explore alternative content within each step (\texttt{d}), or add their own (\texttt{e}). To prevent information overload, the interface keeps some content (\texttt{f}), described in Figure \ref{fig:structured_ui_3}, hidden by default.}
    \Description{}
    \label{fig:structured_ui_all}
\end{figure*}

The \textsc{structured} approach enables using LLMs in a step-by-step reframing process (here, we follow Dorst's nine-step reframing process \cite{dorst:book:reframing}).
Following this approach, designers can use LLMs to generate not only problem frames but also content in each reframing step. When compared to merely seeing alternative frames, this technique might show value in helping designers deepen their understanding of the problem and refine their thought process.
Accordingly, we designed a system wherein LLMs take the content generated in the previous step as input for the next one (this is presented in Figure \ref{fig:structured_diagram}).
We implemented an interface wherein LLM-generated contents are shown in each window, grouped by similarity of purpose in problem reframing (see Figure \ref{fig:structured_ui_all}).
In a similarity to the \textsc{direct} approach, users can click the ``Reframe'' button to prompt the LLMs to go through the entire set of steps or use a refresh button offered in each content window to prompt the LLMs to generate local content.
We designed step-specific prompts for this as shown in tables \ref{tab:strcutured_1}, \ref{tab:strcutured_2}, \ref{tab:strcutured_3}, and \ref{tab:strcutured_4}. Overall, we designed these to enable a prompt-chaining technique \cite{promptChaining:wu:2022}, whereby the outputs can be taken as input to the next prompt.
Our design decisions on the prompts are shown in the corresponding tables.

\begin{figure*}[t]
    \centering
    \small
    \includegraphics[width=\linewidth]{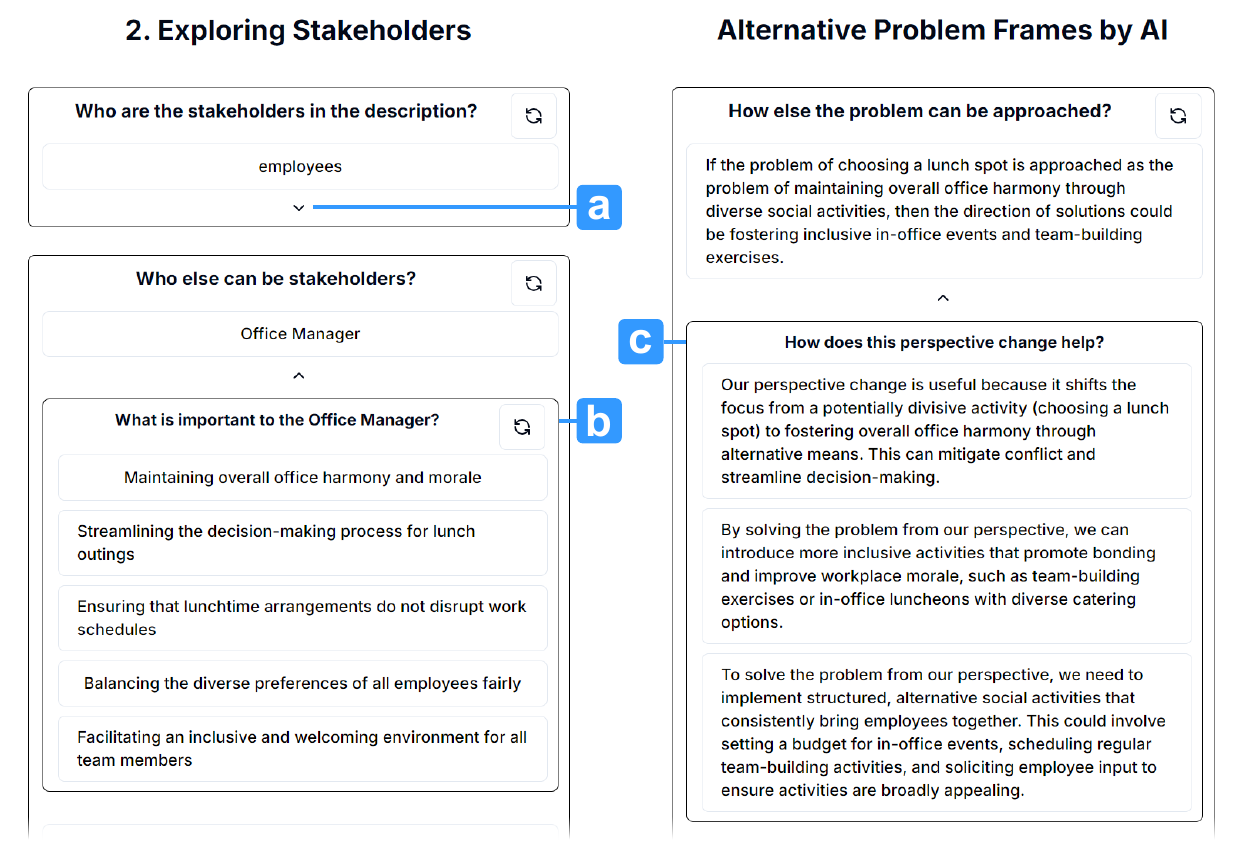}
    \caption{The steps related to the stakeholders and problem frames have hidden windows that fill out the Dorst process. Users can reveal them by clicking the down-arrow button (\texttt{a}). For the former, the window shows what each stakeholder finds important in the problem situation (\texttt{b}). For each problem frame, the window describes how a shift in perspective to the new frame might help address the problem (\texttt{c}); this covers the frame's utility, additional benefits that the frame could bring, and concomitant changes that are feasible.}
    \Description{}
    \label{fig:structured_ui_3}
\end{figure*}

\clearpage

\begin{table}[h]
\centering
\small
\caption{Characterization of the prompts in the \textsc{structured} approach for destructuring of problems. Step 1 and 2's prompts focus on retrieving factual information from the problem description. While GPT-4o was able to list apparent problems without additional guidance, it frequently generated ineffective solutions when the problem description lacked specific solutions, so we instructed GPT-4o to respond with ``No previous solutions'' in these cases. In step 3, GPT-4o struggled to grasp the concept of paradoxical problems at first and often produced a verbose summary of the initial problem. For correction, we introduced a concise explanation of paradoxical problems based on Dorst's work and set a limit to the output length to assure of clarity.}
\begin{tabularx}{\textwidth}{p{0.25\textwidth}X}
\toprule
\textbf{Step 1: Apparent problems} &
Here is a description of a problem situation: \texttt{\{problem description\}}\newline
List apparent problems that are described in the description.\newline
Show only the result in the following format: \texttt{\{output template\}} \\
\midrule
\textbf{Step 2: Ineffective solutions} &
Here is a description of a problem situation: \texttt{\{problem description\}}\newline
From the description, list previous solutions that did not work.\newline
If there are no previous solutions, reply [``No previous solutions.''].\newline
Show only the result in the following format: \texttt{\{output template\}} \\
\midrule
\textbf{Step 3: Paradoxical problems} &
Here is a description of a situation: \texttt{\{problem description\}}\newline
Here are apparent problems: \texttt{\{apparent problems\}}\newline
Here are previous ineffective solutions: \texttt{\{ineffective solutions\}}\newline
What other problems arise by solving or attempting to solve one problem (i.e., paradox)?\newline
If your explanation is based on your speculation, explicitly express your uncertainty.\newline
Explain it without being verbose.\newline
Explain it in maximum two sentences.\newline
Show only the result in the following format: \texttt{\{output template\}} \\
\bottomrule
\end{tabularx}
\label{tab:strcutured_1}
\end{table}

\begin{table}[h]
\centering
\small
\caption{The \textsc{structured} approach's prompts related to exploring stakeholders. We intentionally avoided using the term ``problem'' when referring to the problem description. This helped guide the LLMs to identify stakeholders within a broader context rather than remain confined to a problematic viewpoint. Step 4 focuses on retrieving factual information about stakeholders explicitly mentioned in the problem description, while steps 5 and 6 exploit the LLMs’ generative capacities to suggest potential stakeholders and what each stakeholder values in the given situation. Particularly for step 6, we updated GPT-4o’s system prompt to produce responses from a specific stakeholder's angle, taking advantage of LLMs’ strength in role-playing \cite{roleplay/Shanahan2023, perttu:chi:2023}. We also limited each stakeholder's perspective to five points, since GPT-4o tends to generate excessively verbose and repetitive lists. To further facilitate generation of unique perspectives specific to the stakeholder in question, we provided a list of other stakeholders, enabling ready differentiation.}
\begin{tabularx}{\textwidth}{p{0.25\textwidth}X}
\toprule
\textbf{Step 4: Current stakeholders} &
Here is a description of a situation: \texttt{\{problem description\}}\newline
List stakeholders, people who are involved in the situation.\newline
List the ones stated in the description.\newline
Show only the result in the following format: \texttt{\{output template\}} \\
\midrule
\textbf{Step 5:Potential stakeholders} &
Here is a description of a situation: \texttt{\{problem description\}}\newline
List potential stakeholders who are not stated in the description.\newline
Show only the result in the following format: \texttt{\{output template\}} \\
\midrule
\textbf{Step 6: Stakeholders'\newline perspectives} &
You are a \texttt{``\{stakeholder\}''} in the following situation: \texttt{\{problem description\}}\newline
Answer my question as a \texttt{``\{stakeholder\}''}.\newline
In contrast to other stakeholders such as \texttt{\{list of other stakeholders\}}, what are the things that only you care about?\newline
Tell me five fundamentally different ones.\newline
Show only the result in the following format: \texttt{\{output template\}} \\
\bottomrule
\end{tabularx}
\label{tab:strcutured_2}
\end{table}

\begin{table}[h]
\centering
\small
\caption{How we implemented the \textsc{structured} approach particularly for ``Synthesizing Problem Themes.'' We experimented with two prompting strategies: a) clustering stakeholders' perspectives, then generating a theme for each cluster and b) identifying $N$ unique needs from stakeholders' perspectives. In testing with GPT-4o, we found that the clustering often failed to capture stakeholders' views accurately. It tended to exclude several perspectives (seemingly at random), group unrelated topics together, or generate overly simplistic themes (e.g., producing vague themes such as ``Inclusiveness'' instead of more specific ones such as ``Facilitating an inclusive and welcoming environment''). In contrast, prompting GPT-4o with a specific number of unique needs to identify resulted in more distinct perspectives. We set the number of unique needs to 10 and allowed users to (re-)generate alternative themes if they deemed this necessary.}
\begin{tabularx}{\textwidth}{p{0.25\textwidth}X}
\toprule
\textbf{Step 7: Themes} &
Here are what stakeholders care about in a problem situation: \texttt{\{stakeholders' perspectives\}}\newline
Capture the ten most fundamental and unique needs.\newline
Write each need briefly, without being verbose.\newline
Show only the result in the following format: \texttt{\{output template\}} \\
\bottomrule
\end{tabularx}
\label{tab:strcutured_3}
\end{table}

\begin{table}[h]
\centering
\small
\caption{Our system's handling of steps 8 and 9, which reverses the order from Dorst’s procedures of having designers generate problem frames and then assess them. In our system, LLMs evaluate how viewing the problem in terms of each theme could assist in reframing (step 8), then use the assessments to generate an alternative problem frame (step 9). While designers typically make assumptions about alternative perspectives before generating frames, LLMs do not naturally follow such a process. Changing the order exploits LLMs’ reasoning capabilities to guide more effective reframing. For the assessment, we applied Dorst’s evaluation criteria for problem frames \cite{dorst:book:reframing} to guarantee high-quality evaluation in reframing.}
\begin{tabularx}{\textwidth}{p{0.25\textwidth}X}
\toprule
\textbf{Step 8: Frame assessment} &
Here are descriptions of a problem: \texttt{\{apparent problems\}}\newline
Here is why solving the problem is difficult: \texttt{\{paradoxical problems\}}\newline
To overcome the difficulty of solving the problem, we now look at the problem from a completely different angle, ``a problem of \texttt{\{theme\}}''.\newline
Think step by step to explain the benefits of our perspective change in detail:\newline
- Usefulness: How useful is our perspective change for solving the difficult problem?\newline
- Additional benefits: What are the other benefits of solving the problem from our perspective?\newline
- Feasibility: What changes need to be made to solve the problem from our perspective?\newline
Show only the result in the following format: \texttt{\{output template\}} \\
\midrule
\textbf{Step 9: Problem frames} &
Here is a description of a problem: \texttt{\{problem description\}}\newline
Here is why solving the problem is difficult: \texttt{\{paradoxical problems\}}\newline
In response, we now look at the problem from a completely different angle, ``a problem of \texttt{\{theme\}}''.\newline
Here is how we expect the perspective change to solve the problem: \texttt{\{frame assessment\}}\newline
According to our expectation, reframe the problem in a single sentence.\newline
Write it concisely without being verbose.\newline
Show only the result in the following format: \texttt{\{problem frame template\}} \\
\bottomrule
\end{tabularx}
\label{tab:strcutured_4}
\end{table}

\clearpage

\section{Quiz for Testing Participants' Understanding of Problem Reframing}
\label{appendix:quiz}

We designed a quiz to test participants' competence in problem reframing. This is reproduced as Table \ref{tab:quiz}. While designers are often categorized as ``experts'' on the basis of their number of years of experience in the industry, experience on its own is not an accurate proxy for how well designers understand the concept of problem reframing \cite{cross:2004:design_expertise, ericsson:2018:experience_expertise}. For instance, some designers with several years of practice under their belt might not have engaged in problem reframing or may have an incorrect understanding of the process, due to a lack of formal training or to exposure to poor definitions. We developed five quiz items that helped us assess participants' knowledge spanning the key concepts, processes, and outcomes of problem reframing. The questions and response options were based on the core literature about problem reframing \cite{schon:book:reflective, dorst:book:reframing, crilly:2021:co-evolution}. So that participants could not easily get categorized as experts by choosing responses at random. we utilized items that could have multiple correct answers.

\begin{table}[h]
\centering
\caption{The quiz items used in our study and the correct answers}
\label{tab:quiz}
\begin{tabularx}{\textwidth}{Xc}
\toprule
\textbf{Q1. Which are \textit{incorrect} descriptions of problem reframing? Select all.} &  \\ 
\midrule
It requires iterative exploration of the problem and solutions space. & \xmark \\ 
It is exploring alternative ways of approaching the problem. & \xmark \\ 
It often requires collecting more information about stakeholders. & \xmark \\ 
It is developing solutions to initial problems only. & \cmark \\ 
\midrule
\textbf{Q2. Which are the tasks in problem reframing? Select all.} & \\ 
\midrule
Understanding why the problem is difficult to solve. & \cmark \\ 
Assessing the feasibility of potential solutions to the reframed problems. & \cmark \\ 
Gathering fund for implementing solutions. & \xmark \\ 
Filtering out stakeholders who are causing the problem. & \xmark \\ 
\midrule
\textbf{Q3. Which cases describe reframing design problems? Select all.} & \\ 
\midrule
I reframed a math problem to find a better solution. & \xmark \\ 
I redefined a problem based on what stakeholders find important to themselves. & \cmark \\ 
I summarized the original problem description into a problem framing. & \xmark \\ 
I defined a new problem by uncovering the root cause of the problem. & \xmark \\ 
\midrule
\textbf{Q4. Which are the final outcomes of reframing a problem? Select all.} & \\ 
\midrule
Improved understanding of the problem. & \cmark \\ 
A description of the final problem framing. & \cmark \\ 
A consensus on the most effective solution. & \xmark \\ 
A finalized prototype of the solution. & \xmark \\ 
\midrule
\textbf{Q5. Which are \textit{not} important for problem reframing? Select all.} & \\ 
\midrule
Overcoming fixation on the initial problem. & \xmark \\ 
Blaming stakeholders who are causing the problem. & \cmark \\ 
Exploring potential solutions to the reframed problems. & \xmark \\ 
Speculating additional benefits of solutions other than solving the initial problem. & \xmark \\ 
\bottomrule
\end{tabularx}
\end{table}

\rrSection{
\section{Three Design Problems for Reframing}
\label{appendix:problems}

We articulated three design problems for reframing, described below.

\begin{enumerate}
    \item \textbf{Aging}: As life expectancy increases, more elderly individuals are working longer, particularly in the delivery industry. However, their age-related declines in reaction time and vision have led to an increased rate of traffic accidents involving senior drivers. This led to public concern and pressure on the government to enhance safety on the roads. In response, the government has created a law to limit the maximum age for drivers in the delivery industry. On the contrary, this solution has sparked a backlash from senior drivers, who feel that the law unfairly limits their ability to work and maintain their livelihood.
    \item \textbf{Misbehavior}: A public healthcare service is struggling with the littering of cigarette butts. Despite having multiple smoking booths with ashtrays, smokers tend to throw their cigarette butts on the street. In response, the healthcare service promoted anti-smoking campaigns and increased the fines for littering. However, the littering problems have not been solved, leading to frustration among pedestrians and residents. Furthermore, the idea of prohibiting smoking has caused strong opposition from smokers, and non-smokers worry that prohibition will push smokers to find hidden places for smoking, exacerbating the littering problem.
    \item \textbf{Preference conflict}: In an office, employees have established a tradition of having lunch together, which has strengthened their team bond and promoted social interactions. However, since the number of employees has increased, choosing a lunch spot that can satisfy everyone has become quite challenging. Now, employees face multiple conflicting preferences, such as food quality over price or distance to restaurants and vice versa. These considerations often lead to having an excessively long discussion or everyone having lunch separately, undermining their valued lunchtime culture.
\end{enumerate}
}

\rrSection{
\section{Pilot study}
\label{appendix:pilot}
We conducted a pilot study to confirm the feasibility of reframing the design problems we had prepared (reproduced in Appendix \ref{appendix:problems}). Using the same procedure and platforms as for the main study, we recruited 36 people (17 self-identifying as women and 19 as men) from design-related industries and randomly assigned three participants to each design problem and approach (i.e., each problem was reframed by 12 participants).

With the pilot study, we internally reviewed the frames' quality.
Overall, the participants could generate alternative frames.
For each problem, we found 2--3 of the 12 problem frames novel or useful: they discussed perspectives that were new / not included in the original problem descriptions. For example, P19 reframed the problem of cigarette-butt litter as one of fostering an aesthetically pleasing environment that can autonomously deter misbehaviors rather than punish them. In the case of road safety declining because of society aging, P6 suggested focusing on designing alternative career-development programs for seniors instead of merely banning them from driving. Likewise, P17 introduced the perspective of transforming lunch-spot conflicts into social activities that can enhance group culture. There were also a few low-quality framings, aligned with the original problem descriptions, such as punishing misbehavior / using coercive techniques to prevent senior citizens from driving or resolving conflicts. 
The pilot study showed that our design problems can be reframed into both low- and high-quality problem frames. This supports concluding that the problems are challenging yet able to be reframed.
}


\rrSection{
\section{Participants' Demographics}
\label{appendix:demographic}
}

\begin{table}[ht]
\centering
\small
\caption{Participant demographics connected with each number of correct answers in the quiz}
\begin{tabularx}{0.8\textwidth}{c c c c c}
\toprule
\textbf{Number of} & \textbf{Age} & \textbf{Self-identified gender}: & \textbf{Design experience} & \textbf{$N$} \\
\textbf{correct answers} & mean (and SD) & woman, man, other & mean years (SD) & \\
\midrule
0   & 31.97 (9.06)  & 19, 15, 0  & 5.62 (4.57) &  34 \\
1   & 34.88 (11.19)  & 38, 42, 1  & 7.78 (7.21) &  81 \\
2   & 35.66 (11.00)  & 50, 33, 3  & 8.88 (7.83) &  86 \\
3   & 33.20 (10.00)  & 24, 21, 1  & 7.33 (5.83) &  46 \\
4   & 31.14 (10.51)  & 12, 14, 2  & 7.86 (7.68) &  28 \\
5   & 30.40 (3.65)  & 2, 3, 0  & 3.00 (2.35) &  5 \\
\bottomrule
\end{tabularx}
\label{tab:demographics_all}
\end{table}

\begin{table}[ht]
\centering
\small
\caption{Demographics of the experts who evaluated the problem frames}
\begin{tabularx}{0.8\textwidth}{c c c c c}
\toprule
\textbf{Number of} & \textbf{Age} & \textbf{Self-identified gender}: & \textbf{Design experience} & \textbf{$N$} \\
\textbf{correct answers} & mean (and SD) & woman, man, other & mean years (SD) & \\
\midrule
3   & 32.21 (9.67)  & 1, 2, 0  & 6.33 (3.73) &  3 \\
4   & 32.60 (11.12)  & 4, 6, 1  & 7.86 (7.68) &  11 \\
5   & 32.00 (N/A)  & 0, 1, 0  & 1.00 (N/A) &  1 \\
\bottomrule
\end{tabularx}
\label{tab:demographics_expert}
\end{table}


\end{document}